\documentclass[manuscript, screen, nonacm]{acmart}

\usepackage{multirow}
\usepackage{multicol}
\usepackage{color}
\usepackage{colortbl}
\usepackage{stfloats}
\usepackage{subcaption}
\usepackage{enumitem}
\setlist{nolistsep}
\usepackage{graphicx}
\usepackage{subcaption}
\usepackage{tabularx}
\usepackage{longfbox}
\usepackage{kotex}
\usepackage{xcolor}
\usepackage{tcolorbox}
\usepackage{soul}
\usepackage{makecell}
\usepackage{tikz}
\usepackage{float}
\usepackage{xspace}
\makeatletter
\newdimen\@tempdimd
\makeatother

\newfboxstyle{boxparam}{padding=1.5pt, padding-left=2pt, padding-right=2pt, height=7.5pt, background-color=lightgray, border-radius=2pt, border-right-width=1pt, border-bottom-width=1pt, border-color=gray}

\definecolor{quotebackground}{HTML}{EFEFEF}
\definecolor{tableheader}{HTML}{EFEFEF}
\definecolor{tablegrayline}{HTML}{e0e0e0}

\newcommand{\eg}{\textit{e.g.}}
\newcommand{\ie}{\textit{i.e.}}

\newcommand{\apparatus}{SAL Logger}

\definecolor{bookcolor}{HTML}{C6E4AF} 
\definecolor{webcolor}{HTML}{B8D0E3}
\definecolor{chatbotcolor}{HTML}{EBC6C4} 

\newfboxstyle{toolboxstyle}{
  padding=2.5pt,
  padding-left=2.5pt,
  padding-right=2.5pt,
  border-style=none,
  height=7pt,
  border-radius=3pt
}
\newcommand{\Book}{\lfbox[toolboxstyle, background-color=bookcolor]{\sffamily\textcolor{black}{Book}}}
\newcommand{\Web}{\lfbox[toolboxstyle, background-color=webcolor]{\sffamily\textcolor{black}{Search Engine}}}
\newcommand{\Chatbot}{\lfbox[toolboxstyle, background-color=chatbotcolor]{\sffamily\textcolor{black}{Chatbot}}}

\definecolor{deepnavy}{HTML}{1B2A4E} 

\newtcbox{\tagbox}{%
  on line,
  colback=deepnavy,
  coltext=white,
  boxrule=0pt,
  arc=0pt,
  boxsep=1pt,
  left=2pt,
  right=2pt,
  top=1pt,
  bottom=1pt,
  height=2.5ex,        
  valign=center        
}

\definecolor{lilac}{RGB}{150, 100, 220}

\newboolean{clean}
\setboolean{clean}{true} 

\ifthenelse{\boolean{clean}}{%
  \newcommand{\revised}[1]{#1}
  
}{%
  \newcommand{\revised}[1]{\textcolor{red}{#1}}
  
}

\newcommand{\idx}[1]{}

\newboolean{cameraclean}
\setboolean{cameraclean}{true} 

\newcommand{\cameraready}[1]{\ifthenelse{\boolean{cameraclean}}{#1}{\textcolor{blue}{#1}}}

\newcommand{\deleted}[1]{%
  \ifthenelse{\boolean{clean}}{}{%
    \textcolor{cyan}{\st{#1}}%
  }%
}

\newcommand{\deletedsubsection}[1]{%
  \ifthenelse{\boolean{clean}}{}{%
    \subsection{\textcolor{cyan}{[Deleted] #1}}%
  }%
}

\newcommand{\ipstart}[1]{\vspace{1mm} \noindent{\textbf{\textit{#1.}}}}

\usepackage{makecell} 
\usepackage{multirow} 
\newcolumntype{C}[1]{>{\centering\arraybackslash}m{#1}}
\newcolumntype{L}[1]{>{\raggedright\arraybackslash}m{#1}}

\AtBeginDocument{%
  }

\setcopyright{acmlicensed}
\copyrightyear{2018}
\acmYear{2018}
\acmDOI{XXXXXXX.XXXXXXX}
\acmConference[Conference acronym 'XX]{Make sure to enter the correct
  conference title from your rights confirmation email}{June 03--05,
  2018}{Woodstock, NY}
\acmISBN{978-1-4503-XXXX-X/2018/06}

\acmSubmissionID{6386}
\begin{document}

\title{Easy Come, Easy Go? Examining the Perceptions and Learning Effects of LLM-based Chatbot in the Context of Search-as-Learning}
\renewcommand{\shorttitle}{Examining the Perception and Learning Effects of LLM-based Chatbot in the Context of Search-as-Learning}


\author{Yeonsun Yang}
\orcid{0009-0002-4653-8159}
\affiliation{%
  \institution{Electrical Engineering and Computer Science, DGIST}
  \city{Daegu}
  \country{Republic of Korea}}
\email{diddustjs98@dgist.ac.kr}

\author{Ahyeon Shin}
\orcid{0009-0009-6587-8002}
\affiliation{%
  \institution{Electrical Engineering and Computer Science, DGIST}
  \city{Daegu}
  \country{Republic of Korea}}
\email{ahyeon@dgist.ac.kr}

\author{Mincheol Kang}
\orcid{0009-0006-5907-7733}
\affiliation{%
  \institution{Electrical Engineering and Computer Science, DGIST}
  \city{Daegu}
  \country{Republic of Korea}}
\email{presidentmc@dgist.ac.kr}

\author{Jiheon Kang}
\orcid{0009-0003-6187-3171}
\affiliation{%
  \institution{Electrical Engineering and Computer Science, DGIST}
  \city{Daegu}
  \country{Republic of Korea}}
\email{kangjiheon@dgist.ac.kr}

\author{Xu Wang}
\orcid{0000-0001-5551-0815}
\affiliation{%
  \institution{Computer Science and Engineering, University of Michigan}
  \city{Ann Arbor}
  \state{Michigan}
  \country{USA}}
\email{xwanghci@umich.edu}

\author{Jean Y. Song}
\orcid{0000-0003-4379-3971}
\affiliation{%
  \institution{Humanities, Arts, and Social Sciences, Yonsei University}
  \city{Incheon}
  \country{Republic of Korea}}
\email{jeansong@yonsei.ac.kr}
\begin{abstract}

The cognitive process of Search-as-Learning (SAL) is most effective when searching promotes active encoding of information. The rise of LLMs-based chatbots, which provide instant answers, introduces a trade-off between efficiency and depth of processing. Such answer-centric approaches accelerate information access, but they also raise concerns about shallower learning. To examine these issues in the context of SAL, we conducted a large-scale survey of educators and students to capture perceived risks and benefits of LLM-based chatbots. In addition, we adopted the encoding-storage paradigm to design a within-subjects experiment, where participants (N=92) engaged in SAL tasks using three different modalities: books, search engines, and chatbots. 
\revised{Our findings provide a counterintuitive insight into stakeholder concerns: while LLM-based chatbots and search engines validated perceived benefits on learning efficiency by outperforming book-based search in immediate conceptual understanding, they did not result in a long-term inferiority as feared. Our study provides insights for designing human-AI collaborative learning systems that promote cognitive engagement by balancing learning efficiency and long-term knowledge retention.}


\end{abstract}

\renewcommand{\shortauthors}{Yang et al.}





\maketitle

\section{INTRODUCTION}

Search-as-Learning (SAL)~\cite{sal1} represents a powerful method for knowledge acquisition when harnessed effectively. During an active SAL process, learners engage in the natural flow of information processing~\cite{lindsay2013human}, as they move through the critical stages of encoding, storage, and retrieval to build a complete mental model of a concept~\cite{estes2022handbook,simon1978information,feigenbaum1959information}. Traditionally, this process has relied on conventional resources such as web search engines, academic databases, libraries, and static materials like textbooks~\cite{jansen2009using,weyer1982design,kim2012understanding}. These established modalities often require significant cognitive effort, as learners must actively structure and synthesize information on their own~\cite{vakkari2016searching}. For example, a learner using a textbook for SAL must actively navigate the table of contents and index, cross-reference concepts, and synthesize information from different chapters to form a coherent understanding. Similarly, when using web search engines, the learner must evaluate the credibility of multiple sources, compare conflicting information, and organize disparate pieces of text, images, and videos.


In contrast to these traditional methods, recent advances in Large Language Models (LLMs) and conversational chatbots introduce a fundamentally different learning experience~\cite{kazemitabaar2024codeaid,perez2020rediscovering}. Rather than merely retrieving lists of information based on keywords, these tools act as collaborative assistants~\cite{mo2025conversational}, delivering synthesized summaries and engaging in dialogue. The shift to conversational chatbots makes the search easier and more ``answer-centric,'' introducing a fundamental trade-off: higher efficiency in information access versus the potential for shallower knowledge gain. 
Although learners can achieve higher information throughput~\cite{r2006throughput} by acquiring more facts per unit of time, this efficiency may come at the expense of long-term learning retention~\cite{joiner2008long} and the formation of robust knowledge structures~\cite{novak1990concept}. 


Despite the promise~\cite{pardos2024chatgpt,peters2024ai} and accompanying concerns~\cite{lee2025impact,kosmyna2025your} about the use of LLM-based tools in education, there remains a lack of empirical research that systematically evaluates their effectiveness in the context of SAL. We believe that this gap exists because the roles of LLM-based chatbots---including how they are perceived by both instructors and learners and their actual impact on learning outcomes---have not been adequately compared to traditional SAL tools like textbooks and search engines. 
To address this, our research focuses on a systematic investigation of LLM-based chatbots as a new tool for SAL. By specifically comparing their effects with those of traditional tools, we aim to uncover the unique contributions and drawbacks of chatbots and investigate how their role might differ from that of conventional methods. We expect that this comparative approach can offer crucial insight into how LLMs can best support SAL practices and inform the design and development of effective learning strategies for human-AI collaborative mechanisms. 

To this end, we conducted a large-scale investigation comparing three SAL modalities: textbooks, a web search engine (Google~\cite{google}), and an LLM-based chatbot (ChatGPT~\cite{chatgpt4o}) during July 2024. Through large-scale surveys with educators ($N=75$) and university students ($N=92$), we first provide a comparative overview of the \emph{perceived} potentials and pitfalls of using LLMs for learning. We then complement these findings with a within-subjects experiment inspired by the encoding-storage paradigm~\cite{kiewra1989review}, where the 92 participants engaged in SAL tasks using each of the three modalities in a randomized order. The tasks involved a sequence of activities including information interaction, reviewing, concept map drawing, and closed-book testing. We examined how the three modalities affect learning outcomes in terms of learning efficiency, \revised{information throughput and accuracy, and long-term knowledge retention.} 
The results reveal a trade-off \revised{between immediate learning gain and longer-term retention: while the chatbot condition facilitated the highest immediate recall of concepts and inter-conceptual links, and together with the search engine, achieved significantly higher immediate test scores than the book condition, these benefits were rather temporary. The lack of significant differences on the two-week delayed closed-book tests of the three modalities indicates that this initial performance boost did not translate into superior long-term retention. Further analysis based on Bloom's taxonomy show that the immediate performance benefits of the LLM-based chatbot and search engine were confined to the \emph{Understand} level, with no significant differences observed in the \emph{Apply} or \emph{Analyze} levels compared to text books.}

\revised{Interestingly, our findings challenge the validity of educators' concerns regarding limited cognitive engagement in LLM-based chatbots for SAL. While survey respondents worried that the passive nature of chatbot interaction would compromise the depth of learning and internalization, the comparable long-term retention scores across chatbot, search engine, and book conditions indicate otherwise. The anticipated \emph{negative impact} on knowledge retention was not observed; rather, retention rates were similar across all conditions, even as the chatbot and search engine provided the efficient and adaptive support users expected. Although the initial superiority of digital tools was transient, it did not lead to a knowledge deficit in the long run compared to traditional text book reading. This suggests that while AI and search engines facilitate rapid information intake through lower cognitive load, they do not appear to compromise the ultimate volume of retained knowledge. We discuss whether this ``easy come, easy go'' phenomenon stems from cognitive retention ceilings or the absence of desirable difficulty, highlighting the need for future designs that balance the ease of AI assistance with the cognitive engagement required for consolidation.}



\section{RELATED WORK}

\subsection{Information Processing Theory and SAL} 

The theoretical foundation for our work is rooted in information processing theory~\cite{simon1978information, estes2022handbook}, a cognitive framework that models the human mind as a system for processing, storing, and retrieving information. Information processing theory posits that learning is not a passive event but an active process where learners move information through distinct memory stages: sensory memory, working memory, and long-term memory~\cite{atkinson1968human}. This perspective emphasizes that the depth of learning is directly tied to the level of cognitive effort, or mental investment, required to move and encode information from short-term to long-term memory~\cite{paas1993efficiency}. 

This view of learning is particularly relevant to SAL, which has long been a foundational concept in cognitive science, examining how learners transform information-seeking into knowledge acquisition~\cite{abcd1,abcd2,vakkari2016searching}. 
Historically, this framework was primarily built around traditional tools that provide static information. Early SAL studies explored how learners navigate and synthesize content from compiled resources like textbooks and encyclopedias~\cite{sage2019reading,rockinson2013electronic,edyburn1991fact}. The advent of web search engines (e.g., Google~\cite{google}, Bing~\cite{bing}) marked a significant shift, offering vast but disparate information that required learners to shoulder a substantial cognitive burden for source evaluation and synthesis~\cite{eppler2004concept,thompson2013digital,walraven2008information}. 
Because the act of navigating, evaluating, and integrating disparate information pieces is what gives the SAL process its learning value, a traditional SAL experience aligns with the principles of information processing theory. The inherent difficulty of the task forces the learner to exert the cognitive effort necessary for \idx{15}\revised{higher-order thinking}.

However, the recent emergence of conversational chatbots based on LLMs represents a fundamentally new inflection point, prompting us to reconsider how to best support SAL in this new era of AI. Unlike traditional tools, chatbots serve as collaborative assistants~\cite{kazemitabaar2024codeaid,avula2018searchbots, dhillon2024shaping}, delivering synthesized information and engaging in dialogue. This shifts the learner's role from a passive retriever to an active collaborator, presenting a new set of challenges that extend beyond simple information access. Our study directly addresses this need by empirically comparing the effectiveness of LLM-based chatbots against traditional tools: textbooks and search engines. Through this comparative approach, our goal is to identify the unique benefits and drawbacks of the modalities and ultimately propose design guidelines for AI-mediated learning environments that cultivate effective SAL practices.

\subsection{LLMs in Education: A Dual Perspective on Promise and Peril}
Recent advances in artificial intelligence (AI), particularly LLMs, have led to their increasing use in cognitively challenging tasks. These tools have the potential to significantly enhance productivity in many domains, including writing~\cite{dhillon2024shaping} and other collaborative tasks~\cite{gao2023coaicoder,shaer2024ai}. However, applying this technology to education raises considerable concerns~\cite{lee2025impact,kosmyna2025your}, as the core goal of learning is not just to produce correct answers but to develop critical thinking skills and deeper knowledge. This pursuit of intellectual growth requires significant cognitive engagement, a process that an over-reliance on AI assistance could compromise. In this paper, we investigate this tension in the context of SAL, a primary mode of self-directed knowledge acquisition.

Prior research on the use of LLMs in education has presented a dual narrative of promise and peril. On one hand, studies have demonstrated the potential of LLMs to act as personalized tutors~\cite{chen2023gptutor,pal2024autotutor}, generate customized learning content~\cite{draxler2023relevance}, and provide instant feedback~\cite{phung2023generating}, leading to increased efficiency and motivation~\cite{leong2024putting}. These applications suggest that LLMs could streamline the learning process and make it more accessible. On the other hand, a growing body of work has raised critical concerns. They highlight risks such as the potential for LLMs to generate inaccurate or hallucinated information, which can mislead learners and undermine the quality of knowledge acquisition~\cite{pardos2024chatgpt,bastani2024generative}. Additionally, researchers have voiced concerns that over-reliance on LLMs may impede the development of critical thinking, problem-solving, and information literacy skills that are central to \idx{15}\revised{higher-order thinking}~\cite{peters2024ai,dwivedi2023opinion,mogavi2024chatgpt,weidinger2022taxonomy}. While these studies provide a valuable conceptual foundation for understanding the opportunities and risks, they often lack a direct, empirical comparison of how LLMs' unique capabilities and features translate into concrete learning outcomes compared to different SAL tools.

Our study directly addresses these research gaps by employing a large-scale within-subject design that compares three distinct SAL tools, integrating both student and educator perceptions with empirical metrics of efficiency, accuracy, and long-term retention to offer a comprehensive understanding of \idx{15}\revised{LLM-based chatbots for SAL}.

\section{STUDY}

\begin{figure}[ht]
  \centering
  \includegraphics[width=\linewidth]{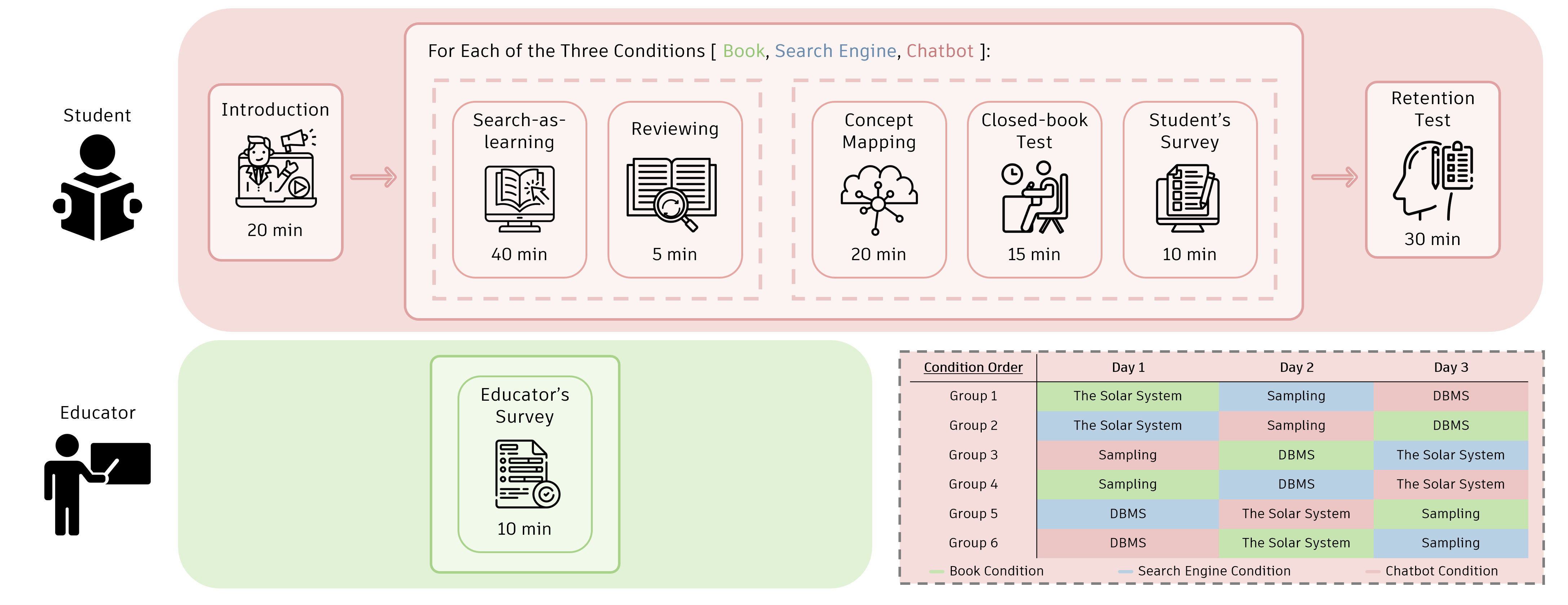}
  \caption{Procedure over the five steps in a mixed-method within-subject study with 92 university students.}
  \Description{Figure 1 represents the overall timeline of the study for students and educators. Upper one shows students' timeline, which presents the subsequent events- Introduction, Search-as-learning, Reviewing, Concept Mapping, Closed-book test, Student's Survey, Retention Test. The lower one shows that the educator conducted a survey for 10 minutes. Additionally, the figure illustrates the condition order that student went through during  their search-as-learning phase.}
  \label{procedure}
\end{figure}

\begin{figure}[h]
  \centering
  \includegraphics[width=\linewidth]{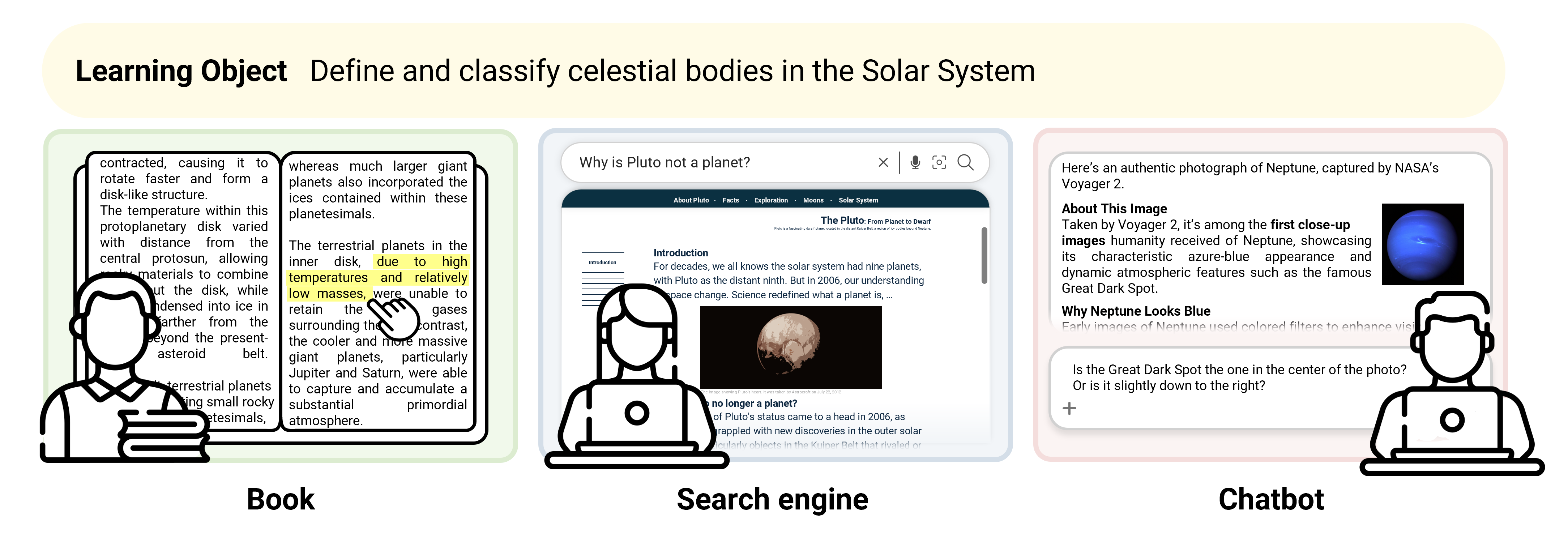}
  \caption{Despite a shared learning objective, students' engagement patterns, interaction types, and the resultant cognitive burdens can diverge significantly across learning tools. By comparatively analyzing different SAL tools: books, search engines, and chatbots, we aim to clarify chatbots' unique contributions and limitations, offering insights into better LLM support in SAL practices. (The images are from actual experiment logs: book search highlight of S88, search engine query of S69, and chatbot log of S36.)
  }
  \Description{This figure shows the study screens of students learning about the solar system with a book, a search engine, and a chatbot, respectively. The student using the book is reading by dragging a given sentence. The student using the search engine is searching for why Pluto is not a planet. The student using the chatbot is asking the chatbot about the location of the Great Dark Spot on the image of Neptune it provided.}
  \label{teaser}
\end{figure}

To systematically investigate the effectiveness of LLM-based chatbots for SAL, we adopt a mixed-methods approach that structures the exploration of both perceived and empirical effects (\autoref{teaser}). 
Specifically, we surveyed educators ($N=75$; E1--E75) and students ($N=92$; S1--S92) to gain a comprehensive understanding of both pedagogical (teaching) and practical (learning) perspectives. 
In addition, we conducted a controlled within-subject user study with students ($N=92$) to provide empirical evidence on learning outcomes, focusing on search efficiency, accuracy of acquired knowledge, and delayed retention.
To draw complementary insights, we conducted the student survey after the user study, where they were asked to reflect on their learning experience with each condition. 
The educator survey was conducted concurrently with the user study to gather their perspectives.
Note that the surveys and experiment were conducted during July 2024.
\idx{7}\revised{The study was approved by the Institutional Review Board at the author's university.}

Our research questions are as follows:

\begin{itemize}
    \item RQ1: [Perception] What are the main benefits and risks perceived by students and educators when using LLM-based chatbots for SAL?
    \item RQ2: [Efficiency]  How do the completion time and the number of \revised{search Q\&A pairs} differ between using LLM-based chatbots and traditional search methods?
    \item RQ3: [Accuracy] How does the \revised{information throughput and} accuracy of knowledge acquired with LLM-based chatbots compare to that acquired with traditional search methods?
    \item RQ4: [Retention] How does the long-term retention of knowledge acquired with LLM-based chatbots compare to that acquired with traditional search methods?
    
\end{itemize}


\subsection{Survey on Perceptions of SAL Tools: Textbooks, Web Search Engines, and LLM-based Chatbots}
\ipstart{Survey with Educators}
We recruited 75 South Korean educators with varying educational backgrounds and experiences through snowball sampling and online advertising. All educators work in educational institutes in South Korea, except for two who are based in the United States. Our respondents included 33\% females, 64\% males, and 3\% who did not disclose their gender. The age distribution was as follows: 25\% were 20–29 years old, 24\% were 30–39 years old, 29\% were 40–49 years old, and 21\% were 50 years or older. Our sample consisted mainly of university or high school teachers (93\% of the sample). Most of them either work in the STEM field (Science, Technology, Engineering, Mathematics; 57\%) or in the HSS field (Humanities and Social Sciences; 33\%) and 63\% of them had experience using LLM-based chatbots. The survey took about 10 minutes to complete. 
\idx{7}\revised{Educator participation was voluntary, with no monetary compensation. }

In the survey, open-ended questions were asked to explore their perspectives on the benefits and drawbacks of LLM-based chatbots in the SAL process, \eg{}, their thoughts on the benefits and risks of using LLM-based chatbots. 
They were further asked to rank the three tools in terms of which they would encourage students to use when studying unfamiliar concepts independently in courses, along with their rationale. 
We used a rank instead of a Likert scale to make direct comparisons between tools, revealing their relative preferences and priorities, and to avoid the tendency to give neutral answers~\cite{yannakakis2015ratings,harzing2009rating}. 


\ipstart{Survey with Students}
In a post-survey after the experiment, students were asked to reflect on their SAL experiences with books, search engines, and an LLM-based chatbots. The survey included open-ended questions for comparative evaluations of the three tools, their intended future use in university coursework, and the perceived benefits and risks of each tool with supporting rationales. The survey took about 10 minutes to complete. \revised{Compensation for the survey was included in the total payment for the entire experiment.}

\subsection{Experiment Design}
We performed a within-subject experiment with 92 university students, where they engaged in self-directed learning to achieve predefined learning objectives in the three SAL conditions.
The study was conducted in a university-level STEM context, given that STEM texts typically involve complex structures and technical terminology that place high demands on learners’ cognitive resources.
To allow direct comparison across conditions, we designed the study in a web-based environment.
As the experimental apparatus, we developed a browser extension to log search throughput (\ie{}, the number of \revised{search Q\&A pairs} per unit time). 
We introduced learning tasks to assess immediate outcomes and delayed retention in learning.
\idx{3,8}\revised{To minimize fatigue and order effects, we used a counterbalanced design in which the three modalities and three STEM modules were rotated across participants, resulting in six sequence groups (\autoref{procedure}).}

\begin{table}[ht]
\renewcommand{\arraystretch}{1.3}
\caption{Comparison of SAL conditions across four dimensions.}
\resizebox{\textwidth}{!}{%
\begin{tabular}{l@{\hspace{1cm}}c@{\hspace{1.5cm}}c@{\hspace{1.5cm}}c@{\hspace{1.5cm}}c}
\toprule
\textbf{Condition} & \begin{tabular}[x]{@{}c@{}}\textbf{Information} \\ \textbf{Throughput} \end{tabular} & 
\begin{tabular}[x]{@{}c@{}}\textbf{Learner's structuring} \\ \textbf{and encoding burden} \end{tabular}
& \textbf{Information quality} 
& \textbf{External storage quality} \\

\midrule
\Book{}    & Low    & High     & High (reliable, structured) & Low--Medium (self-made) \\
\Web{}  & Medium & Medium & Low-High (uncertified) & Medium \\
\Chatbot{} & High   & Low   & Low-High (hallucination) & High (pre-synthesized) \\
\bottomrule
\end{tabular}%
}
\label{conditions}
\Description{This table presents comparison of SAL conditions across four dimensions(Information Throughput, Learner's structuring and encoding burden, Information quality, External storage quality). While a chatbot offers high information throughput and a low burden on the learner, it poses a risk to information quality. Conversely, a book guarantees high information quality but at the cost of lower throughput and a higher learner burden. The web generally represents a middle ground between these two extremes.}
\end{table}

\subsubsection{Searh-as-Learning Conditions}
\label{sec:conditions}
Inspired by the encoding–storage paradigm~\cite{kiewra1989review, chen2025more}, we designed three experimental conditions (Table~\ref{conditions}). The \Book condition replicates academic textbooks and scholarly publications, which is characterized by its \textbf{low information throughput}. Accessing specific information within a physical book requires a sequential and manual process of locating the text, navigating chapters, and scanning pages ---notably slower than digital alternatives. Despite this, \idx{6}textbooks impose a \textbf{high structuring and encoding burden for learners}~\revised{\cite{kintsch1988burden, sweller1998burden}}. While the content itself is typically highly structured and logically presented by experts, learners must actively engage in deep reading, comprehension, and critical analysis to internalize, summarize, and integrate this knowledge into their own cognitive frameworks. This active processing is crucial for effective learning, but demands significant cognitive effort. Correspondingly, textbooks consistently have \textbf{high information quality}, resulting from rigorous editorial, peer-review, and fact-checking processes that ensure accuracy, reliability, and conceptual coherence. Finally, textbooks typically lead to \textbf{low to medium external storage quality}. Although the physical book serves as a robust external repository, its utility as personal external storage (\eg{} for quick recall or reference) is largely dependent on the learner's own efforts to create supplementary aids such as notes, highlights, or summaries. The effectiveness of this self-constructed external storage varies significantly with the learner's skill and diligence.

The \Web condition offer a \textbf{medium information throughput}, providing rapid access to a vast index of online resources through keywords and query based searches. Queries yield numerous links almost instantaneously, allowing for quick initial retrieval. However, the learner must still manually navigate through various webpages, filter irrelevant results, and critically evaluate the information, which slows down the overall processing compared to direct answers. This process contributes to a \textbf{medium structuring and encoding burden for learners}. Learners are responsible for sifting through fragmented and often conflicting information from diverse sources, assessing credibility, and then synthesizing these disparate pieces into a coherent understanding. Search engine results can vary in \textbf{low to high information quality} because most content are uncertified, where anyone can create content online. This wide range reflects the open nature of the internet. Content can vary from highly reliable, peer-reviewed articles to unsubstantiated claims, or misinformation. The absence of a universal vetting mechanism for all indexed content means that learners must exercise significant discretion and critical evaluation. Consequently, the \textbf{external storage quality is medium}. While learners can leverage browser functionalities (\eg{} bookmarks, tab categorization) or dedicated tools to manage searched pages, the engine itself provides uncurated and disparate information. This requires that the learners personally organize and synthesize this content into genuinely usable and coherent external knowledge.

The \Chatbot condition provides \textbf{high information throughput}, delivering direct, synthesized \emph{answers} and summaries to complex queries almost instantly. This eliminates the need for users to navigate multiple sources or manually filter results, thus maximizing the rate at which relevant information is presented. Consequently, the \textbf{learner's structuring and encoding burden is low}. Chatbots provide information that is largely pre-digested, often structured into coherent paragraphs or lists, which significantly reduces the cognitive effort required for the learner to structure and integrate new knowledge internally. However, the \textbf{information quality spans low to high} mainly due to the hallucination effect. While they can synthesize information into accurate and insightful responses (high quality), they are prone to generating confident yet factually incorrect or entirely fabricated information (low quality). This inherent unpredictability requires careful verification by the user. Finally, the \textbf{external storage quality is high} because chatbots provide information that is already processed, summarized, and often structured in a ready-to-use format. This pre-synthesized knowledge minimizes the learner's effort in finding, organizing, and preparing information, effectively serving as an immediate and accessible extension of their own cognitive resources.

For the experimental setup, we selected a predefined textbook~\cite{carroll2017introduction,2023sampling,2023Introduc} for \Book condition, Google~\cite{google} for \Web condition, and ChatGPT-4o~\cite{chatgpt4o} for \Chatbot condition as representative tools.

\subsubsection{Experimental Apparatus}

\begin{figure}[ht]
  \centering
  \includegraphics[width=0.9\linewidth]{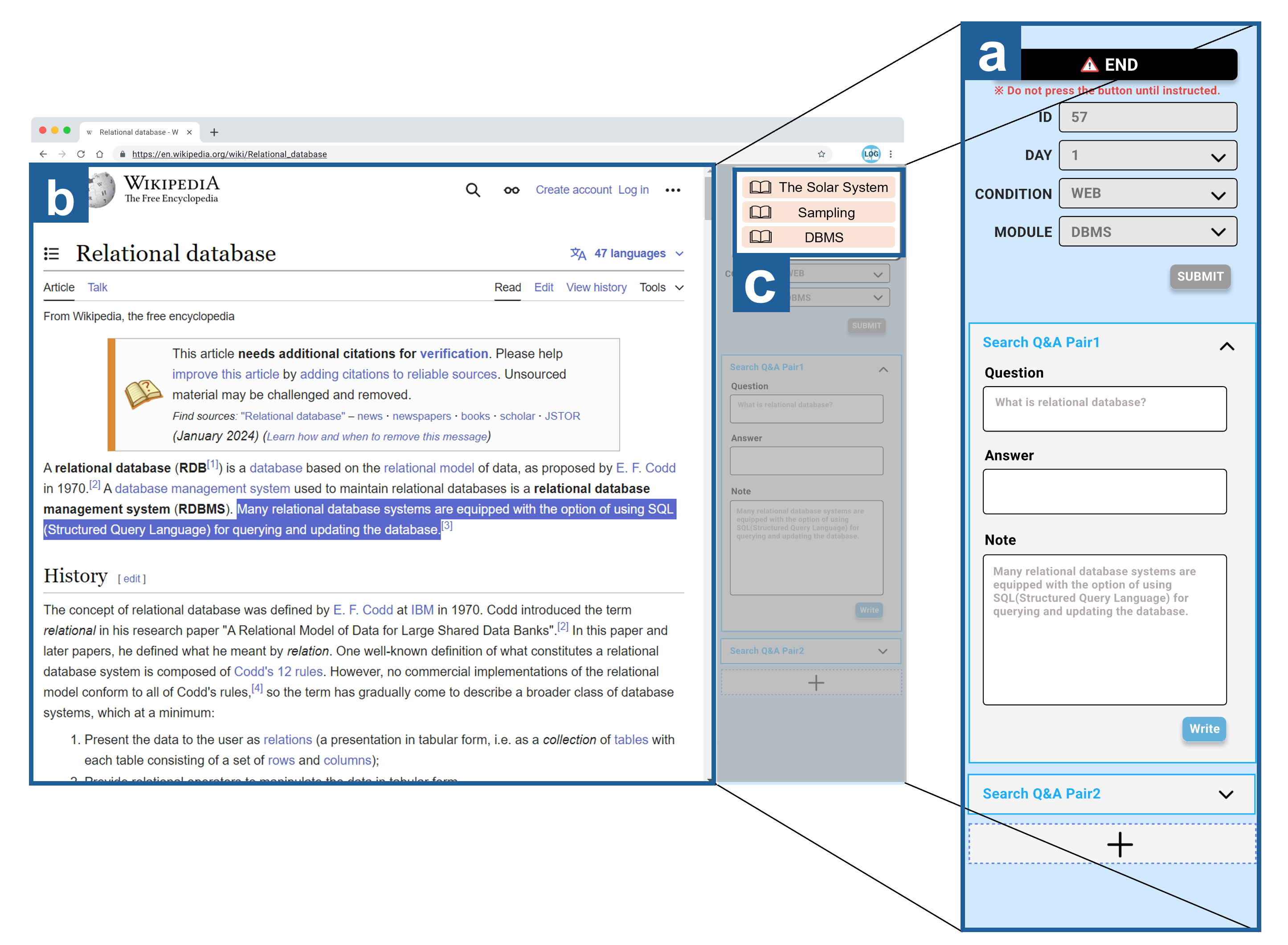}
  \caption{Overview of the SAL logger of \Web. The interfaces for \Book and \Chatbot follow a similar structure, consisting of two interface components: \tagbox{a} the content panel and \tagbox{b} the note panel. A Chrome extension plug-in \tagbox{c} was provided for the \Book condition to access pre-selected textbooks in PDF format. The experiment starts by inputting information \revised{such as} participant ID, session (from Day 1 to Day 3), experiment condition, and study subject on top of the note panel. Participants start by initiating a question themselves in the Question section. As they engage in self-guided search and learning on the content panel, they can drag relevant information or organize it into their own words in the Note section. Once they find an answer to their query, participants complete the Answer section and submit it. They can add new notes by clicking the ’+’ button. 
  }
  \Description{The figure presents an overview of the Search-as-Learning (SAL) logger, featuring two main interface components: (a) the note panel and (b) the browser window. The browser displays a learning module (in this case, DBMS) with relevant content. The note panel allows participants to input their experimental session details, write down their queries, and store the answers and memos. Additionally, (c) shows the Chrome extension plug-in with pre-selected text, Books for each module.
  }
  \label{SAL_logger}
\end{figure}

The \apparatus{} (\autoref{SAL_logger}) runs on Chrome browser and records timestamps of user interactions, including participants' queries and accessed resources. We designed the logger to switch among the three conditions of SAL tools introduced in \autoref{sec:conditions} based on an experiment condition variable. 
The logger consisted of two interface components: a content panel (\autoref{SAL_logger}-\tagbox{a}), where participants interact with the assigned SAL tool to perform tasks, and a note panel on the right (\autoref{SAL_logger}-\tagbox{b}), where they collect and organize information within each SAL search Q\&A pair. 
We developed the SAL logger as a Chrome extension using HTML and JavaScript. It has a client-server architecture that stores search histories, and collects learning logs. A dedicated server was implemented to ensure the security of learning data and fulfill requirements for robust client-server operation within our experimental framework. 
The server is implemented using Node.js and stores data in JSON format.

Our system supports the three experimental conditions as follows:
\begin{itemize}
    \item \Book : The textbooks were sourced from university-level courses and provided in PDF format, covering all predefined learning objectives (The Solar System~\cite{carroll2017introduction}, 568 pages; Sampling~\cite{2023sampling}, 461 pages; DBMS~\cite{2023Introduc}, 228 pages). They were translated into Korean to remove language barriers and integrated into our \apparatus{} (\autoref{SAL_logger}-\tagbox{c}). To distinguish this condition from web documents, participants were instructed to use the table of contents and index instead of full-text search, reflecting the high encoding effort of book-based search.
    \item \Web : The browser window displays multiple web documents, blogs, or other resources. We restricted access to Google~\cite{google} to provide a consistent search experience, limited to traditional search results without LLM integration (as of July 2024). To control for personalization due to search history, all searches were conducted using a dedicated experimental account. 
    \item \Chatbot : The content panel displayed the ChatGPT interface for conversational search. We restricted this condition to ChatGPT-4o~\cite{chatgpt4o}, selected for its enhanced speed and multimodal capabilities. To isolate and assess the intrinsic characteristics of chatbot-only interactions, we deactivated the built-in web browsing feature.
\end{itemize}

\subsubsection{Participants and Procedure}
We recruited 92 university students from our institute located in South Korea, through mailing lists and online community advertisements (average age=21, 46 males and 46 females). 
To mitigate potential confounding effects of prior knowledge on study results, we selected STEM modules specifically not covered in our institute's curriculum. Furthermore, a screening process was implemented to actively exclude participants who possessed pre-existing familiarity with the target topics. 
The screening test consisted of three MCQs at the Remember level~\cite{krathwohl2002revision} of Bloom’s taxonomy to assess prior knowledge, as this level requires simple recall of factual information. 
Students who scored full marks (3 out of 3) on the screening test were filtered out. Participants were randomly assigned to one of six possible counterbalanced sequences of conditions (shown in the bottom left box in \autoref{procedure}).  
The average score of the screening test of all study participants was 0.55 ($SD=0.68$), with no significant differences (Kruskal–Wallis $H=0.675$, $p=0.70$) across the condition orders, confirming the successful mitigation of prior knowledge bias across all sequences.

The experiment began with an introduction to the study, providing participants with a thorough explanation of the tasks and procedures. Participants were asked to submit informed consent and received an online link to the \apparatus{}. They then installed the \apparatus{} on the Chrome browser and started studying the predefined learning objectives (Table~\ref{LOs}) using a designated SAL tool. 
During the 40-minute study, participants formulated their own questions in the note panel whenever they identified knowledge gaps or had internal questions. To address these questions, they searched for information using the SAL tool and organized key insights or pertinent content in the Note section. Once questions were resolved, participants submitted their \revised{search Q\&A pairs} with answers through the system. 
After completing the SAL task, they were given five minutes to review and reflect on their \revised{search Q\&A pairs} on the \apparatus. 

Then they completed a 20-minute concept map drawing task without access to any learning materials, hand-drawing labeled nodes and linking them to illustrate relationships among concepts.
\idx{4}\revised{Participants were instructed to construct dense and comprehensive concept maps.}
This task was included to capture the effective information throughput achieved by each modaliity. 
After completing the concept map drawing task, participants conducted a closed-book test, where they had 15 minutes to solve nine MCQs. Finally, participants completed a survey via an online form, which included questions about their experiences and perceptions of the given SAL condition. 

This process was repeated over \textbf{three consecutive days}, wherein both the study topics and the assigned SAL tools were systematically varied (please refer to \autoref{procedure}). Lastly, two weeks later, participants completed a 30-minute retention test. 
\idx{7}\revised{Participants received 66,000 KRW (approximately 45 USD) as compensation.}

\subsubsection{Learning Tasks and Materials}

\begin{table}[ht]
\renewcommand{\arraystretch}{1.4}
\caption{Nine learning objectives used in the within-subject controlled experiment.}
\resizebox{\textwidth}{!}{%
\begin{tabular}{p{3cm} p{5cm} p{5cm} p{5cm}}
\toprule
\textbf{Bloom's taxonomy} & \textbf{The Solar System} & \textbf{Sampling} & \textbf{DBMS} \\
\midrule
Understand & Define and classify celestial bodies in the Solar System. & Define the concept of sampling. & Define the concepts of databases and tables. \\
\midrule
Apply & Explain and compare the properties and characteristics of planets in the Solar System. & Explain and compare probability sampling and non-probability sampling. & Explain and compare RDBMS and non-RDBMS. \\
\midrule
Analyze & Analyze planetary motion using Kepler's Laws. & Classify various probability sampling techniques and apply their formulas. & Analyze data with CRUD operations using MySQL and basic syntax.\\
\bottomrule
\end{tabular}%
}
\Description{The table shows learning objectives (LOs) for three modules: Astronomy, Sampling, and Database. For Astronomy, LO 1 is "Define and classify planets and dwarf planets," LO 2 is "Explain the properties and characteristics of planets in the Solar System," and LO 3 is "Apply Kepler’s Laws of planetary motion." For Sampling, LO 1 is "Define the concept of sampling," LO 2 is "Explain and compare probability sampling and non-probability sampling," and LO 3 is "Classify various probability sampling techniques and apply their formulas." For Database, LO 1 is "Define the concepts of databases and tables," LO 2 is "Explain and compare RDBMS and non-RDBMS," and LO 3 is "Apply CRUD operations using MySQL with the basic syntax." The main keywords in each learning objective are highlighted in bold.}
\label{LOs}
\end{table}

To control for participants' prior knowledge, we selected STEM modules not included in our institute's curriculum. The authors conducted three iterative rounds of discussion to set three modules from distinct STEM domains: The Solar System (Astronomy), Sampling (Statistics), and Database Management Systems (Computer Science).
Drawing on university-level coursework available from online platforms such as Coursera\footnote{https://www.coursera.org/} and edX\footnote{https://www.edx.org/}, we derived three learning objectives for each module, aligned with the Understand, Apply, and Analyze levels of Bloom’s taxonomy~\cite{bloom1968learning,krathwohl2002revision} (Table~\ref{LOs}). Participants had to study all three learning objectives for each module assigned for the day. 
We excluded higher-level taxonomies because they are difficult to assess within the limited learning timeframe.

\noindent \emph{Concept map:} To capture information throughput, 
we employed Novakian concept mapping~\cite{canas2004varieties}, a method commonly used to capture learners’ mental models by representing networks of connections between related concepts~\cite{novak1990concept,delugach2016knowledge,moon2018case,o2004measuring,reese2004assessment}. 

\noindent \emph{Multiple-choice questions (MCQs):} The post-test consisted of nine MCQs designed to align with the predefined learning objectives. 
\idx{5}\revised{The MCQs were first generated through few-shot prompting, and the final items were selected and refined through question-level evaluation (\autoref{tab:question-level}) involving four of the authors.
To ensure the validity of the MCQs, three domain experts were recruited to conduct a quiz-level evaluation using three metrics (\autoref{tab:expert_eval}).
The expert ratings indicated acceptable quality: The Solar System scored (2,3,3), and both Sampling and DBMS each scored (3,2,3) for structure, redundancy, and usefulness.}
We provide details on question generation, quality evaluation, and example MCQs in~\autoref{sec:MCQgeneration}.
To further assess long-term retention, we administered a delayed test two weeks later using the same questions as the post-test, with randomized order and options.

\begin{table}[ht]
\renewcommand{\arraystretch}{1.3}
\caption{The quiz-level evaluation metric was used by subject matter experts to finalize the MCQ set. Each question was assessed based on three criteria: Structure, Redundancy, and Usefulness.}
\centering
\resizebox{0.8\textwidth}{!}{%
\begin{tabular}{p{3cm} p{9cm} p{3cm}}
\toprule
\textbf{Metric} & \textbf{Definition} & \textbf{Evaluation} \\
\midrule
Structure & It measures whether the set of questions makes sense together. & Ordinal metric (1–3) \\
\midrule
Redundancy & It measures if there is redundancy/repetition within the quiz. & Ordinal metric (1–3) \\
\midrule
Usefulness & It measures if a teacher would use the quiz in an assessment they create for their own class. & Ordinal metric (1–4) \\
\bottomrule
\end{tabular}%
}
\Description{The table presents three quiz-level evaluation metrics: Structure, Redundancy, and Usefulness. Structure evaluates whether the set of questions makes sense together. Redundancy assesses repetition within the quiz. Usefulness measures whether a teacher would include the quiz in an actual assessment. Each metric is evaluated on an ordinal scale.}
\label{tab:expert_eval}
\end{table}

\section{Data Analysis Measurement}
\subsection{Survey Analysis}
To compare preferences for SAL tools across educators and students, we performed Friedman's test~\cite{friedman1937use}. For post-hoc analysis, we employed Conover tests~\cite{conover1979multiple} with Bonferroni correction to avoid potential multiple comparison problems. The mean rank metric~\cite{alvo2014statistical} was used to calculate the average rank, ensuring robustness for non-parametric statistical tests. In addition to analyzing perceptions of LLM-based \idx{15}\revised{chatbots for} SAL compared to conventional tools, we conducted theoretical coding~\cite{muller2012grounded} on open-ended responses and comments to uncover the underlying reasons behind the tool preferences. Two of the authors reviewed the responses and classified them into perceived benefits and risks of adopting LLMs in the SAL process. Conflicts were resolved through iterative discussions, and the two authors achieved inter-rater reliability of Krippendorff's alpha of 0.89.

\subsection{Quantitative Outcome Measures}

\subsubsection{Search and Interaction Logs}
To evaluate search efficiency, we analyzed the quantity and duration of \revised{search Q\&A pairs} logged through \apparatus{}. We note that a completed \revised{search Q\&A pair} is a unit in which a single question is issued, relevant information is integrated, and an answer is formulated---representing one cycle of SAL. All timestamps were logged for each event. For comparison, one-way ANOVA tests were conducted across conditions to compare the number of \revised{search Q\&A pairs} and the time spent per \revised{search Q\&A pair}. 

\subsubsection{Closed-book Concept Map Drawing Task}
To assess \idx{4}\revised{information throughput}, we analyzed students' concept maps using count-based network metrics---the number of nodes and the number of edges---which are commonly used in prior literature to estimate individual learners' understanding~\revised{\cite{cromley2024three,besterfield2004scoring,watson2016assessing,collective}}.
To apply these metrics, we reviewed all 276 maps from the participants and resolved synonym conflicts (\eg{}, ``DB'' and ``Database'') through iterative discussions among four of the authors. We then conducted a comparative analysis of the concept map metrics using ANOVA tests. 

\subsubsection{Closed-book and Retention Tests}
To assess the accurate encoding and stable storage of essential knowledge aligned with our learning objectives, we analyzed two types of tests: a closed-book test administered immediately after each condition and a retention test conducted two weeks later. Each test was scored on a 0–9 scale (because nine MCQs were provided per test), and the results were statistically analyzed using one-way ANOVA tests to compare performance across conditions.
\idx{1,2}\revised{To further analyze learning outcomes by cognitive level, we conducted one-way ANOVA tests for the three Bloom levels (Understand, Apply, Analyze) across conditions.
To evaluate knowledge retention, we conducted an ANCOVA on the retention-test scores, with condition as the between-subjects factor and the closed-book test score as the covariate.
}

\revised{\section{\idx{15}FINDINGS}}

\subsection{RQ1: What are the main benefits and risks perceived by students and educators when using LLM-based chatbots for SAL?}
Through our survey analysis of students and educators, we examined their preferences among three SAL tools and further investigated their perceptions of LLM-based chatbots. Our findings reveal a divergence in SAL tool preferences between educators and students, with educators favoring established tools while students show a notable preference for LLM-based chatbots. This disparity stems from a fundamental pedagogical tension: students prioritize efficient, streamlined information encoding for immediate needs, whereas educators emphasize the critical role of cognitive effort and metacognition for deep, durable learning, despite both groups acknowledging the general benefits and risks of LLM-based SAL.


\subsubsection{Preferences}
\begin{figure}[t]
  \centering
  \includegraphics[width=0.7\linewidth]{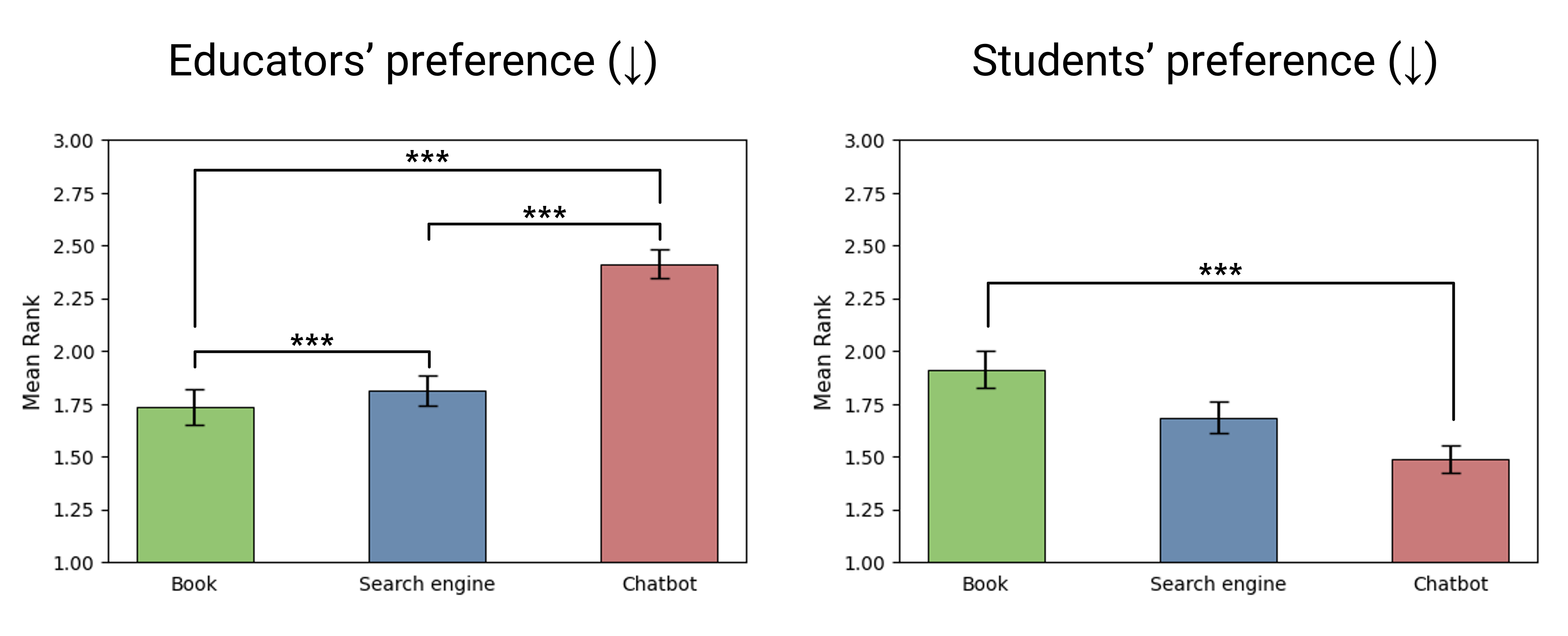}
  \caption{Bar chart showing mean preference ranks from educator and student survey responses. A lower mean rank (↓) indicates higher preference (i.e., 1 = most preferred). The left plot (a) presents educators’ rankings of the three SAL tools, and the right plot (b) shows students’ rankings. Significance is \revised{based on a Friedman test and} marked as $p$ < .001(***).}
  \Description{The figure contains two bar graphs comparing mean preference ranking of each learning sources from (a)educators and (b) students. Graph (a)Educators shows educators preferred Books the most(lowest mean rank), followed by Search engines, and LLM-based chatbots as the least least prefered source. The differences in mean rank between all three sources were statistically significant (p<0.001). Graph (b)Students shows students preferred LLM-based chatbots the most followed by Search engines, and Books. The differences in mean rank between only Books and LLM-based chatbot was statistically significant(p<0.001).}
  \label{preferences}
\end{figure}

As shown in \autoref{preferences}, \idx{12}\revised{a Friedman test revealed} a statistically significant difference in rankings across the three SAL tools ($X^2(2) = 21.62$, $p < .001$).
Post-hoc analysis showed that all three SAL tools differed significantly from one another.
The majority of educators preferred books (52\%; 39 of 75) and search engines (36\%; 27 of 75) as search-as-learning tools, while only 12\% (9 of 75) chose LLM-based chatbots their favorite.
This suggests that educators remain cautious about adopting LLM-based chatbots as SAL tools, showing a clear preference for more established conventional tools. 
Meanwhile, students exhibited an opposite preference order to educators. Notably, the difference in preference between books and LLM-based chatbots was statistically significant ($p<.001$). 

Overall, our comparison highlights a marked difference in tool preferences between students and educators.

\begin{table}[t]
\centering
\renewcommand{\arraystretch}{1.3}
\caption{Perceived benefits and risks of LLM-based chatbots as SAL tools, \protect\revised{reported by educators ($N=75$) and students ($N=92$). Numbers indicate how many respondents mentioned each theme.}}
\resizebox{\textwidth}{!}{%
\begin{tabular}{C{1.6cm}C{4.1cm}L{9cm}C{1.7cm}C{1.7cm}}
\hline\hline
\parbox[c][1cm][c]{\linewidth}{\centering \textbf{Category}} &
  \parbox[c][1cm][c]{\linewidth}{\centering \textbf{Theme}} &
  \parbox[c][1cm][c]{\linewidth}{\centering \textbf{Explanation}} &
  \parbox[c][1cm][c]{\linewidth}{\centering \textbf{Educators}} &
  \parbox[c][1cm][c]{\linewidth}{\centering \textbf{Students}} \\ \hline\hline
\multirow{8}{*}{Benefits} &
  Personalized Learning &
  LLMs can provide adaptive explanations and feedback tailored to learners’ individual needs and progress. &
  \revised{26} &
  \revised{13} \\ \cline{2-5} 
                       & Serendipitous Discovery      & LLMs can provide unexpected but relevant information beyond learners’ initial queries.                 & \revised{8} & \revised{5} \\ \cline{2-5} 
 &
  Diverse Explanations &
  LLMs can present the same concept through multiple explanatory styles, such as simplified summaries, detailed breakdowns, and a variety of materials. &
  \revised{23} &
  \revised{29} \\ \cline{2-5} 
                       & Time Efficiency              & LLMs can provide rapid search and collect extensive information in a short time.                       & \revised{50} & \revised{36} \\ \cline{2-5} 
                       & Selective Learning           & LLMs can provide focused access to desired information without irrelevant content.                     & \revised{0}  & \revised{14} \\ \hline\hline
                       
\multirow{6}{*}{Risks} & Lack of Reliability          & LLMs can provide incorrect or fabricated information without clear sources.                             & \revised{19} & \revised{39} \\ \cline{2-5} 
                       & Reduced Metacognition        & LLMs can reduce metacognitive ability to recognize and reflect on knowledge gaps.                      & \revised{23} & \revised{0}  \\ \cline{2-5} 
 &
  Limited Cognitive Engagement &
  LLMs can hinder learners’ direct engagement in critically evaluating, integrating, and internalizing information. &
  \revised{37} & \revised{0}
   \\ \cline{2-5} 
                       & Prior Knowledge Dependency & LLMs can provide surface-level information when learners lack sufficient prior knowledge on the topic. & \revised{0}   & \revised{24} \\ \hline\hline
\end{tabular}%
}
\Description{The table provides a summary of survey responses from students and educators, categorizing their opinions on LLMs into two main sections: Benefits and Risks. The 'Benefits' category includes five themes: Personalized Learning, Serendipitous Discovery, Diverse Explanations, Time Efficiency, Selective Learning. While all students reported every benefit, educators identified every theme except for Selective Learning. The 'Risks' category details four potential issues: Lack of Reliability, Reduced Metacognition, Limited Cognitive Engagement, Limited Depth of Information. Educators raised the risks of Lack of Reliability, Reduced Metacognition, and Limited Cognitive Engagement. Students acknowledged the risks of Lack of Reliability and Limited Depth of Information.}
\label{risk_and_benefit}
\end{table}

\subsubsection{Perceived Benefits and Risks}
To gain a deeper understanding of the reasons behind these tool preferences, we qualitatively analyzed survey responses on perceived benefits and risks (see Table~\ref{risk_and_benefit}). 
While tool preferences diverged between educators and students, both groups consistently recognized most of the benefits and risks of LLM-based SAL.

In terms of benefits, both students and educators commonly acknowledged the advantages of using LLMs for SAL.
These advantages collectively illustrate a key trade-off we highlight in this work---efficient information encoding---LLMs' ability to rapidly deliver pre-synthesized and structurally organized knowledge tailored to learners' goals, thereby reducing cognitive effort during the encoding phase of learning.
A distinct benefit raised exclusively by students was \emph{selective learning}, which provided focused access to the desired information they are looking for. 
As S5 noted, ``\textit{One of the strengths of using LLMs was that I could quickly focus on the parts I didn’t know, rather than reviewing everything.}''
S9 also remarked, ``\textit{When I searched with Google, I had to sift through tons of information to find what I needed. With ChatGPT, I could directly learn only what I was curious about, which made the process much clearer and less overwhelming.}''
This preference reflects a desire for streamlined and targeted learning experiences, driven by the time constraints students face in balancing academic and personal demands.

In contrast, educators valued more effortful engagement with learning materials, viewing cognitive effort as essential for deep understanding and retention. 
They highlighted risks such as \emph{reduced metacognition} and \emph{limited cognitive engagement}, warning that over-reliance on chatbots could hinder learners from critically evaluating, synthesizing, and internalizing knowledge. 
As E1 emphasized, ``\textit{Even when students are given the same information, they develop creativity and critical thinking by making sense of it in their own way, rather than passively accepting it.}'', and E8 remarked ``\textit{It may take more time, but I believe that deeply focusing on and accurately understanding core concepts is more important than quickly learning superficial facts.}''
However, none of the students raised this issue as a potential risk.
Instead, students pointed to more immediate challenges, such as \emph{prior knowledge dependency}, where difficulty guiding the chatbot arose without sufficient initial understanding. 
Nonetheless, all participants acknowledged that LLMs could be unreliable at times and emphasized the need to interpret their outputs with caution.

Taken together, these findings illustrate a pedagogical tension between short-term encoding and long-term storage. While students prioritized immediate access and ease of use, educators emphasized the importance of cognitive effort and reflective thinking in building durable knowledge.

\subsection{RQ2: How do the completion time and the number of \revised{search Q\&A pairs} differ between using LLM-based chatbots and traditional search methods?} 
\begin{figure}[ht]
  \centering
  \includegraphics[width=0.8\linewidth]{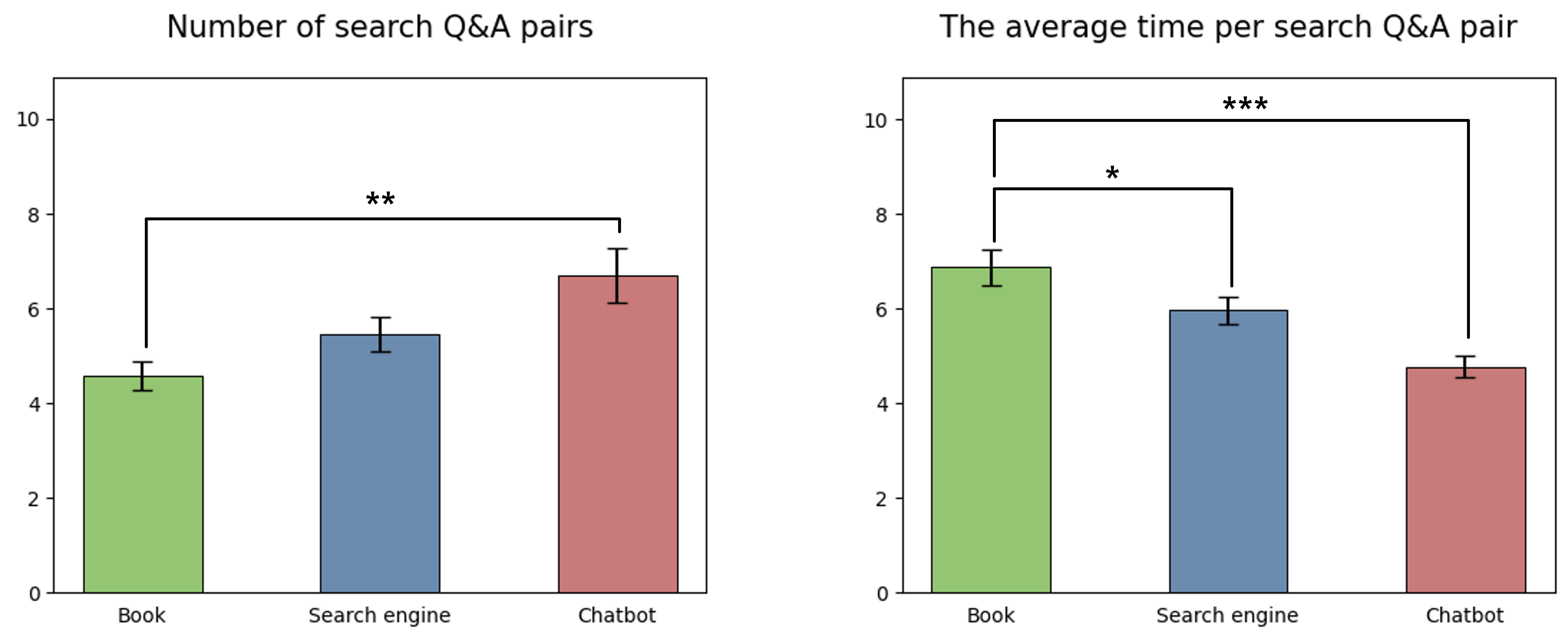}
  \caption{\revised{Bar chart showing the number of search Q\&A pairs (left) and the average time spent per Q\&A pair (right). ANOVA was conducted to compare information throughput across the conditions. Significance is marked as $p$ < 0.1 ($^{+}$), $p$ < 0.05 (*), $p$ < 0.01 (**), or $p$ < 0.001 (***).}}
  \Description{The figure displays two graphs regarding the efficiency of different information sources in completing Q\&A notes. The left graph illustrates the number of search Q\&A pairs. It shows that the largest number of Q\&A pairs were recorded in the chatbot condition, followed by the search engine and then the book. The right graph displays the average time per search Q\&A pair. The average search time was longest in the book condition, followed sequentially by the search engine and the chatbot.}
  \label{QApair}
\end{figure}

During the 40-minute search-as-learning, students completed a total of 442 \revised{search Q\&A pairs} with books, 502 \revised{search Q\&A pairs} with search engines, and 614 \revised{search Q\&A pairs} with LLM-based chatbots \revised{(\autoref{QApair})}.
On average, the number of \revised{search Q\&A pairs} per student was 4.59 ($SD = 2.91$) in the \Book{} condition (the least), 5.45 ($SD = 3.52$) in the \Web{} condition, and 6.69 ($SD = 5.47$) in the \Chatbot{} condition (the most). 
Students completed significantly more \revised{search Q\&A pairs} with LLM-based chatbots than with books ($p < .002$).
The average time per \revised{search Q\&A pair} (in minutes) was longest for books ($M = 6.87, SD = 6.57$), followed by search engines ($M = 5.95, SD = 5.58$), and LLM-based chatbots ($M = 4.77, SD = 4.97$).
Students spent significantly less time per \revised{search Q\&A pair} in the \Chatbot{} condition than in the \Book{} ($p < .001$) and \Web{} ($p = .05$) conditions.

Collectively, these findings suggest that LLM-based chatbots facilitate faster and more frequent information retrieval, thereby empirically substantiating students' perceived efficiency as SAL tools and offering a potential explanation for their strong preference for LLMs over conventional tools.


\begin{figure}[ht]
  \centering
  \includegraphics[width=\linewidth]{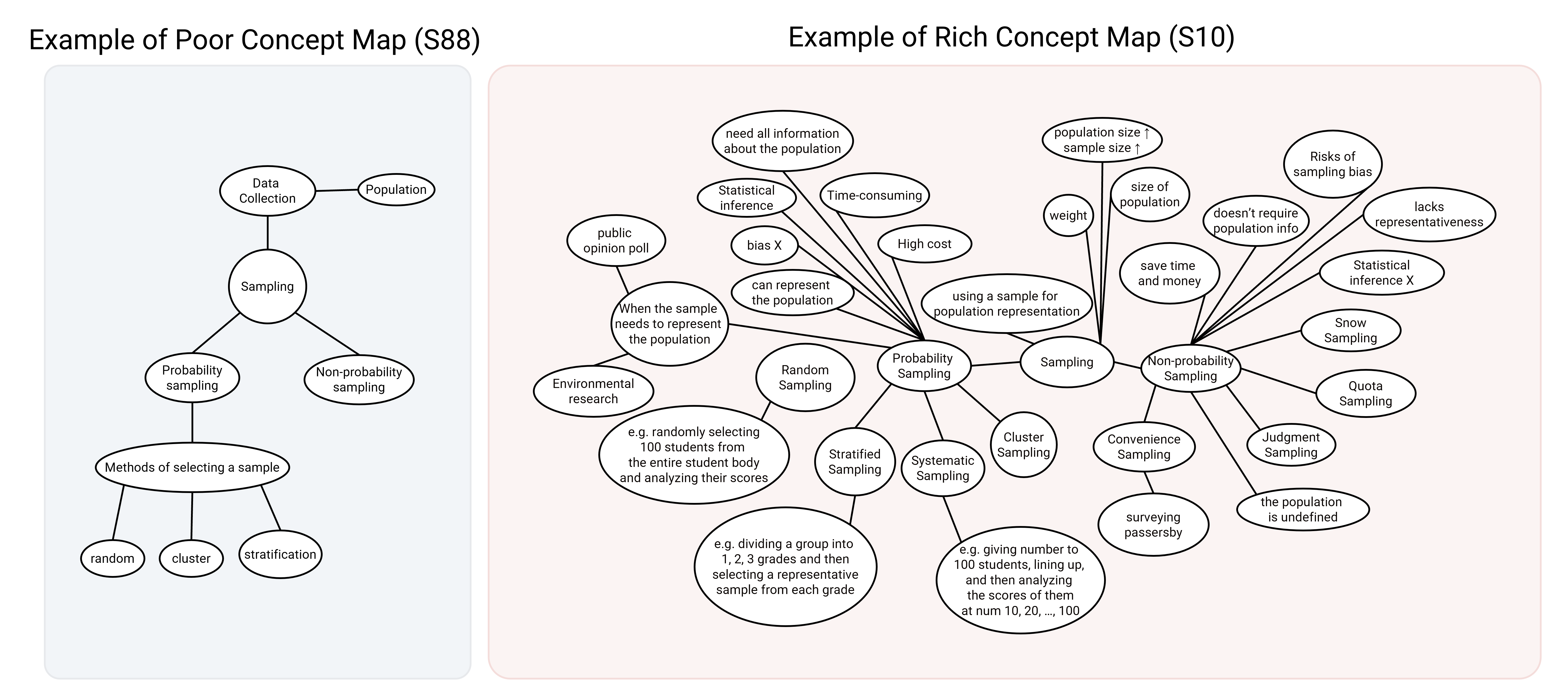}
  \caption{Example of concept maps drawn. The left one has less information compared to the right one. Both are created for the Sampling module (left: search engine condition (S92), right: chatbot condition (S10)). This shows the version before resolving synonym conflicts.}
  \Description{Two examples of concept maps created by different students are shown. The one on the left is a poorly constructed concept map, which is relatively short and lacks sufficient concepts. The one on the right is a well-developed concept map, containing many concepts and a more structured organization.}
  \label{conceptmap}
\end{figure}

\begin{table}[t]
\renewcommand{\arraystretch}{1.3}
\centering
\caption{Summary of quantitative results for RQ3. The left column shows p-values obtained via ANOVA tests for each measurement. The right column shows pairs of conditions where the effect is statistically significant or marginally significant. Significance is marked as $p$ < 0.1 ($^{+}$), $p$ < 0.05 (*), $p$ < 0.01 (**), or $p$ < 0.001 (***).}
\resizebox{\textwidth}{!}{%
\begin{tabular}{C{5.5cm}C{2cm}C{2cm}C{2cm}C{3.5cm}}
\Xhline{1pt}
\parbox[c][6mm][c]{\linewidth}{\centering \textbf{Measurement}} &
\parbox[c][6mm][c]{\linewidth}{\centering \textbf{Book}} &
\parbox[c][6mm][c]{\linewidth}{\centering \textbf{Search Engine}} &
\parbox[c][6mm][c]{\linewidth}{\centering \textbf{Chatbot}} &
\parbox[c][6mm][c]{\linewidth}{\centering \textbf{Post-Hoc Analysis}} \\
\hline
\parbox[c][9mm][c]{\linewidth}{\centering \textbf{Number of Nodes}\\ F(2,272)=7.20, $p<.001$***} &
\parbox[c][9mm][c]{\linewidth}{\centering Mean = 18.67\\ SD = 7.71} &
\parbox[c][9mm][c]{\linewidth}{\centering Mean = 21.21\\ SD = 9.34} &
\parbox[c][9mm][c]{\linewidth}{\centering Mean = 23.76\\ SD = 10.00} &
\parbox[c][9mm][c]{\linewidth}{\centering Chatbot $>$ Book ***} \\
\hline
\parbox[c][9mm][c]{\linewidth}{\centering \textbf{Number of Edges}\\ F(2,272)=4.13, $p<.05$*} &
\parbox[c][9mm][c]{\linewidth}{\centering Mean = 25.67\\ SD = 15.42} &
\parbox[c][9mm][c]{\linewidth}{\centering Mean = 30.64\\ SD = 18.54} &
\parbox[c][9mm][c]{\linewidth}{\centering Mean = 32.83\\ SD = 17.45} &
\parbox[c][9mm][c]{\linewidth}{\centering Chatbot $>$ Book *} \\
\hline
\parbox[c][9mm][c]{\linewidth}{\centering \textbf{Immediate Closed-book Test Score}\\ F(2,272)=9.59, $p<.001$***} &
\parbox[c][9mm][c]{\linewidth}{\centering Mean = 4.83\\ SD = 1.67} &
\parbox[c][9mm][c]{\linewidth}{\centering Mean = 5.55\\ SD = 1.27} &
\parbox[c][9mm][c]{\linewidth}{\centering Mean = 5.80\\ SD = 1.55} &
\parbox[c][9mm][c]{\linewidth}{\centering Chatbot $>$ Book *** \\ Search Engine $>$ Book **} \\
\Xhline{1pt}
\end{tabular}%
}
\Description{This table presents a quantitative summary of a study comparing three conditions (book, search engine, and chatbot) across three measurements (number of nodes, number of edges, and immediate closed-book test score). The results of ANOVA tests and post-hoc analyses show that learning with a chatbot and a search engine led to significantly better outcomes than learning with a book.}
\label{shortterm}
\end{table}

\subsection{RQ3: How does the \revised{information throughput and }accuracy of knowledge acquired with LLM-based chatbots compare to that acquired with traditional search methods?}
To evaluate immediate knowledge acquisition, we examined a total of 276 concept maps created after the 40-minute study phase, focusing on structural metrics such as the number of nodes and edges. 
Across all concept maps, the number of nodes ranged from 7 to 54, and the number of edges ranged from 2 to 122.
\autoref{conceptmap} presents examples of digitized concept maps created by students during the study. 
As shown in \autoref{shortterm}, the average number of nodes was highest in the \Chatbot{} condition ($M = 23.76, SD =10.00$), followed by the \Web{} condition ($M = 21.21, SD =9.34$) and then the \Book{} condition ($M = 18.67, SD =7.71$). 
Similarly, the average number of edges showed the same pattern, with the \Chatbot{} condition including the largest number of edges ($M = 32.83, SD =17.45$), followed by the \Web{} condition ($M = 30.64, SD =18.54$) and the \Book{} condition ($M = 25.67, SD =15.42$). 
A significant difference was observed between the \Chatbot{} and \Book{} conditions for both nodes ($p<.001$) and edges ($p<.05$). 
We found that this difference aligns with the earlier finding that students completed more \revised{search Q\&A pairs} in the \Chatbot{} condition.


To further evaluate the accuracy of acquired knowledge, we examined the colsed-book test scores administered right after the concept map drawing (see last row of \autoref{shortterm}). 
The average scores (scored on a 0 to 9 scale, because there were nine MCQs per condition) were highest in the \Chatbot{} condition ($M=5.80, SD=1.55$), followed by the \Web{} condition ($M=5.55, SD=1.27$), and the \Book{} condition ($M=4.83, SD=1.67$). 
Scores in the \Book{} condition were significantly lower than in both the \Chatbot{} ($p < .001$) and \Web{} conditions ($p < .01$), reflecting the limited search throughput within the given time constraint. 

\idx{1,2}\revised{To examine whether learning gains differed across levels of cognitive processing, we analyzed the closed-book scores by Bloom’s taxonomy—Understand, Apply, and Analyze—and compared performance across conditions at each level~\autoref{tab:mcq}.
Although the overall average scores (scored on 0–3 scale) were highest in the \Chatbot{} condition, followed by \Web{} and \Book{}, modality effects were only significant at the Understand level (Chatbot>Book: $p<.001$; Web>Book: $p<.05$). For the Apply and Analyze levels, no significant differences were observed across conditions.
}

Overall, we observed that the \Chatbot{} condition outperformed the conventional tool conditions on immediate learning outcomes, which contrasts with the perceived risk of the unreliability of LLM outputs.
Rather, these empirical results provide strong evidence for students' preference for LLMs as their primary SAL tools in self-directed learning environment.
\idx{1,2}\revised{However, this benefit was observed mainly at the level of recognizing facts and grasping the meanings of key concepts.}


\begin{figure}[t]
  \centering
  \includegraphics[width=0.35\linewidth]{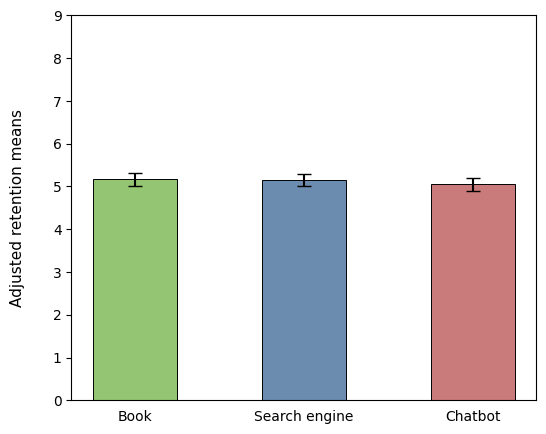}
  \caption{\revised{Adjusted retention means across the conditions, estimated using ANCOVA with the closed-book test score as a covariate.}}
  \Description{The graph shows average score by retention test. This illustrates the mean scores obtained from three different condition(book, search engine, chatbot). The data reveals that the book condition yielded the highest average score(5.164), followed by the search engine(5.153). The chatbot condition produced the lowest average score(5.052) among the three tested conditions.}
  \label{reten_mean}
\end{figure}

\begin{table}[t]
\centering
\caption{\revised{Summary of quantitative results for the closed-book and two-week delayed retention tests across the three Bloom’s taxonomy levels (Understand, Apply, Analyze). Values represent mean scores (0–3 scale per level), with standard deviations shown in parentheses. ANOVAs (closed-book test) and ANCOVAs (retention test) were conducted at each level across conditions. Significance is marked as $p$ < 0.05 (*), or $p$ < 0.001 (***).}}
\resizebox{0.8\textwidth}{!}{%
\begin{tabular}{c|c|c|c|c|c|c}
\Xhline{1pt}
\multicolumn{1}{l|}{} & \multicolumn{3}{c|}{Closed-book test} & \multicolumn{3}{c}{Retention test} \\\cline{2-7}
\multicolumn{1}{l|}{} & Understand        & Apply        & Analyze        & Understand        & Apply       & Analyze       \\\Xhline{1pt}
\textbf{Book}        & 1.58***, *(0.85)        & 1.58(1.03)        & 1.66(0.83)        & 1.62(0.88)        & 1.57(0.98)       & 1.67(0.74)       \\
\textbf{Search engine}         & 1.92*(0.76)        & 1.79(0.92)        & 1.84(0.72)        & 1.89(0.85)        & 1.66(0.93)       & 1.69(0.76)       \\
\textbf{Chatbot}         & 2.07***(0.80)        & 1.84(0.92)        & 1.90(0.77)        & 1.85(0.78)        & 1.66(0.89)       & 1.76(0.78)      \\ \Xhline{1pt}
\end{tabular}%
}
\Description{The table presents the average scores form a Multiple-Choice Question(MCQ) test, differentiating between a Closed-book test and a Retention test, and categorized by information source(Book, Search engine, Chatbot) and cognitive domain(Understanding, Application, Analysis). For the Closed-book test, the Chatbot condition consistently recorded the highest scores across all three cognitive domains. In the subsequent Retention test, this pattern generally persisted, with the Chat group still performing strongly, particularly in Understanding (1.8) and Analysis (1.7). However, the scores across all three sources converged substantially in the Application domain(Book: 1.5, Search engine: 1.6, Chatbot: 1.6).}
\label{tab:mcq}
\end{table}

\subsection{RQ4: How does the long-term retention of knowledge acquired with LLM-based chatbots compare to that acquired with traditional search methods?}
\revised{To examine differences in retention across conditions, we conducted an ANCOVA (\autoref{reten_mean}). The closed-book test score was included as a covariate and significantly predicted retention performance ($F(1,272)=96.78$, $p <.001$), indicating that higher immediate learning gains were associated with higher retention scores. 
After adjusting for the covariate, the adjusted retention scores showed a descriptive pattern (\Book{}:$M=5.16$, \Web{}:$M=5.15$, \Chatbot{}:$M=5.05$), but no significant differences were found across the three conditions.
Similarly, ANCOVA conducted on retention test scores at each cognitive level---Understand, Apply, and Analyze---revealed no significant differences across conditions (\autoref{tab:mcq}).
}


\revised{Taken together, while modality effects were observed in immediate learning performance, such measurable impact was not observed in the retention test scores.
The \Chatbot{} and \Web{} conditions supported more efficient short-term learning outcomes, yet these benefits did not persist into the delayed retention test. Conversely, the \Book{} condition produced smaller initial gains, but retention remained comparable across conditions.}


\section{DISCUSSION}
\subsection{Pedagogical Tension on Efficiency vs. \revised{Knowledge Retention in LLM-Based Chatbots for SAL}}

While both educators and students recognized the trade-offs of LLM use in the SAL contexts, their priorities diverged. Students valued efficient encoding, whereas educators emphasized \revised{internalization of knowledge. 
Our empirical results provide a complicated perspective that includes counterintuitive insight. While LLM-based chatbots facilitated significant short-term gains, these benefits were confined to lower-order thinking and did not translate into superior long-term retention compared to the book condition.} 
We found that LLM-based chatbots enable \revised{ immediate conceptual understanding}, directly supporting students' preference for streamlined and targeted learning experiences.
This pursuit of efficiency is particularly salient given the substantial time demands of STEM curricula, where students often dedicate numerous hours weekly to coursework~\cite{Miller2025WorkloadGap}. Consequently, the ability of \revised{LLM-based chatbots} to rapidly deliver pre-synthesized knowledge becomes an almost indispensable resource, pushing students towards greater reliance on these tools for time-sensitive tasks.

However, despite this perceived efficiency, our within-subjects experiment demonstrated a crucial discrepancy: the short-term gains associated with chatbots did not translate into \revised{higher-order reasoning nor }long-term retention. While the chatbot condition optimized immediate search speed and learning outcomes, allowing students to acquire a larger number of concepts with more inter-conceptual links, \revised{the educational benefits reflected surface-level understanding and tended to decay over time.}
These findings necessitate a cautious approach to integrating LLM-based chatbots into SAL, urging a focus on design strategies that balance efficiency with the imperative for active, effortful, and reflective learning processes to foster robust, long-term knowledge construction.
\subsection{\revised{Interpreting the Cognitive Efficiency and Knowledge Retention across SAL Modalities}}
\revised{Prior work~\cite{Bhattacharya2023longsal,sal1,sal2,sal3,sal4,salreview}} consistently show that active information processing, often associated with a higher cognitive burden, contributes significantly to enhanced self-efficacy, improved learning efficiency, and, crucially, superior long-term memory formation. This phenomenon is often attributed to the principle of ``desired difficulty'', where challenging yet manageable learning tasks lead to deeper processing and more robust knowledge structures \revised{\cite{bjork2011making}}.

The inherent interaction required by traditional books, such as manually navigating tables of contents and indices, cross-referencing concepts, and synthesizing information across chapters, inherently provides this necessary cognitive burden. Unlike the pre-synthesized output of \revised{LLM-based chatbots}, the physical and structured nature of books compels learners to construct their understanding actively, leading to more profound and durable cognitive connections. This active engagement is believed to prevent the passive acceptance of information, instead fostering critical thinking and deeper processing, which are foundational for effective knowledge retention.

Furthermore, the perceived authority and reliability of books likely play a psychological role in this retention advantage. Textbooks, being curated and vetted by experts, carry a strong sense of trustworthiness. This perceived high information quality (as discussed in \autoref{sec:conditions}) likely encourages learners to fully invest in internalizing the content. 
In contrast, LLM-based chatbots, despite their efficiency, carry a known risk of hallucination and generally lack the same authoritative psychological weight. We speculate that this lower perceived authority or potential unreliability of chatbot-generated content may inadvertently prevent learners from fully committing information to long-term memory, as they may unconsciously (or consciously) hold back full trust in the content's veracity. Consequently, learning might remain at a more superficial level, less integrated into \revised{long-term knowledge retention}, due to an underlying lack of absolute confidence in the content's reliability.

\revised{Meanwhile, the web search condition demonstrated comparable search efficiency and short-term learning gains to LLM-based chatbots. Web search requires moderate levels of learner agency, such as iterative query refinement, evaluating source credibility, and triangulating information, which introduces cognitive effort beyond passive response consumption, yet still less than the structured navigation.
Recently, the boundary between web search and LLM-based chatbot interaction is becoming increasingly blurred, as web search engines integrate LLM-powered summarization features, and chatbots increasingly incorporate retrieval-augmented browsing capabilities.
We note that web search was highly preferred by both educators and students in our survey studies, while still producing measurable learning outcomes comparable to those of chatbots in the within-subject experiment. We believe that this positions web search as a promising middle-ground modality. We acknowledge that familiarity may have influenced this performance advantage, as web search represents the tool users reported using most frequently for academic inquiry~\cite{brabazon2016university,rowlands2008google}. Future research is needed to disentangle whether its performance stems from inherent modality affordances versus habitual familiarity.}

\revised{While both web search and chatbot conditions yielded superior short-term learning gains, these advantages did not persist into delayed retention. 
We offer two interpretations of this pattern. 
First, this may reflect cognitive limits in how much pre-synthesized information can be encoded and consolidated, suggesting a natural ceiling on retention capacity~\cite{sweller2011cognitive}. Alternatively, it may indicate that the lack of desirable difficulty constrained deeper processing needed for consolidation. Future work should examine whether the observed decay reflects cognitive limits on retaining rapidly encoded information, and test whether introducing desirable difficulty into LLM-based SAL systems can improve long-term retention.}

\subsection{Towards Designing \revised{Efficient Short-term Gain and Reliable Long-term Retention} using LLMs in SAL
}
Our findings do not suggest abandoning LLM-based tools in SAL. Given their undeniable efficiency and students' strong inclination towards them, the integration of LLMs into educational practices is arguably inevitable. The challenge, therefore, lies not in prohibition but in developing systematic guidelines and design principles that harness their power while mitigating the risks of shallow learning and fostering \revised{long-term retention}. 
A primary design consideration must address the prior knowledge dependency issue raised among students. Current LLMs, by default, lack the sophisticated scaffolding mechanisms common in effective pedagogy. Unlike human instructors who adapt their guidance based on a learner's current understanding, general chatbots do not inherently detect a user's knowledge level to provide tailored, scaffolded instruction. Future SAL-oriented LLM interfaces should incorporate features that proactively assess a learner's existing knowledge, perhaps through initial querying or diagnostic interactions, and then dynamically adapt their responses. This intelligent scaffolding, possibly through advanced prompt engineering, could guide learners through a structured inquiry process, gradually increasing complexity and encouraging deeper engagement rather than simply delivering a final answer through designing ``desired difficulty''. 

Furthermore, our results strongly advocate for blended learning strategies \revised{\cite{graham2006blended}} that intentionally combine the strengths of LLMs with those of traditional tools. While LLMs excel at rapid information throughput and initial encoding, \revised{these benefits did not carry over to long-term retention and higher-order thinking}. Educational designs should, therefore, seek to integrate LLM interactions with more reflective activities. For instance, LLMs could serve as initial information gatherers or summarizers, and then these outputs should be brought into a classroom setting for critical discussion, peer review, or deeper analysis using traditional resources. This ``flipped classroom\revised{\cite{akccayir2018flipped}}'', where AI supports initial content acquisition, could free up valuable in-class time for collaborative problem-solving, debates, and instructor-led activities designed to foster the deeper conceptual understanding and critical thinking necessary for long-term retention.

\subsection{Limitations}~\label{limitation}
Our study has several limitations. First, it was not conducted longitudinally, which may have restricted our ability to capture more natural usage patterns in a real-world educational setting. Second, the controlled experiment was limited to a university-level STEM course with structured, fact-based content, which may have reduced the likelihood of encountering hallucination or misinformation compared to subjects involving complex interpretations (e.g., humanities) or rapidly evolving knowledge domains (e.g., semiconductor processing). Third, the participant pool was restricted to university students, which may limit the generalizability of the findings to learners with different goals or expertise levels. Lastly, while we focused on GPT-based LLM chatbots to examine autonomous learning and information retrieval, future work should assess the performance of LLMs across broader subject areas, include more diverse learner populations, and investigate emerging hybrid search tools that combine LLMs with traditional search engines to evaluate whether they address current limitations.



\section{CONCLUSION}
This research systematically investigated the effectiveness of LLM-based chatbots as a novel Search-as-Learning (SAL) tool, comparing them against traditional methods: textbooks and web search engines. Our large-scale study, involving both educator and student surveys alongside a within-subjects experiment, uncovered a fundamental pedagogical tension: students prioritize the immediate efficiency of LLM-based tools, while educators emphasize the importance of cognitive effort for deeper \revised{learning and knowledge retention}. 
While survey results indicated strong perceived benefits and efficiency from LLM-powered chatbots, particularly among students, our empirical experiment revealed a crucial discrepancy. The chatbot condition indeed facilitated faster learning and allowed students to acquire a larger number of concepts with more inter-conceptual links in the short term. However, this immediate efficiency did not translate into \revised{higher-order thinking or} long-term retention.
These findings underscore a \revised{complex trade-off between processing efficiency and knowledge retention. The observed `easy come, easy go' phenomenon suggests that while AI accelerates information access, it hits a natural ceiling on retention capacity, yielding long-term outcomes comparable to traditional methods. Rather than simply cautioning against AI adoption, future designs should aim to balance the convenience of AI synthesis with he benefits of established tools, strategically optimizing SAL practices for efficient, robust, and effective learning experience.}


\bibliographystyle{ACM-Reference-Format}
\bibliography{reference}

@misc{chatgpt4o,
    title={{ChatGPT-4o}},
    url={https://openai.com/index/hello-gpt-4o},
    author={{OpenAI}},
    year= {2025},
    note={Accessed: Sep 1, 2025}
}

@misc{google,
  author={{Google}},
  title={{Google Search}},
  year={2025},
  url={https://www.google.com/},
  note={Accessed: Sep 1, 2025}
}

@misc{bing,
  author={{Microsoft}},
  title={{Bing Search}},
  year={2025},
  url={https://www.bing.com/},
  note={Accessed: Sep 1, 2025}
}

@inproceedings{bhattacharya2023longsal,
  title={LongSAL: A Longitudinal Search as Learning Study with University Students},
  author={Bhattacharya, Nilavra},
  booktitle={Extended Abstracts of the 2023 CHI Conference on Human Factors in Computing Systems},
  pages={1--8},
  year={2023}
}

@book{abcd1,
  title={Information seeking in electronic environments},
  author={Marchionini, Gary},
  year={1995},
  publisher={Cambridge university press}
}

@article{abcd2,
  title={Study of interactive feedback during mediated information retrieval},
  author={Spink, Amanda},
  journal={Journal of the american society for information science},
  volume={48},
  number={5},
  pages={382--394},
  year={1997},
  publisher={Wiley Online Library}
}

@article{sal1,
  title={Towards searching as a learning process: A review of current perspectives and future directions},
  author={Rieh, Soo Young and Collins-Thompson, Kevyn and Hansen, Preben and Lee, Hye-Jung},
  journal={Journal of Information Science},
  volume={42},
  number={1},
  pages={19--34},
  year={2016},
  publisher={SAGE Publications Sage UK: London, England}
}

@inproceedings{sal2,
  title={Analyzing knowledge gain of users in informational search sessions on the web},
  author={Gadiraju, Ujwal and Yu, Ran and Dietze, Stefan and Holtz, Peter},
  booktitle={Proceedings of the 2018 conference on human information interaction \& retrieval},
  pages={2--11},
  year={2018}
}

@inproceedings{sal3,
  title={The effects of learning objectives on searchers' perceptions and behaviors},
  author={Urgo, Kelsey and Arguello, Jaime and Capra, Robert},
  booktitle={Proceedings of the 2020 acm sigir on international conference on theory of information retrieval},
  pages={77--84},
  year={2020}
}

@article{sal4,
  title={Information searching strategies in web-based science learning: The role of Internet self-efficacy},
  author={Tsai, Meng-Jung and Tsai, Chin-Chung},
  journal={Innovations in education and Teaching International},
  volume={40},
  number={1},
  pages={43--50},
  year={2003},
  publisher={Taylor \& Francis}
}

@article{salreview,
  title={Towards searching as a learning process: A review of current perspectives and future directions},
  author={Rieh, Soo Young and Collins-Thompson, Kevyn and Hansen, Preben and Lee, Hye-Jung},
  journal={Journal of Information Science},
  volume={42},
  number={1},
  pages={19--34},
  year={2016},
  publisher={SAGE Publications Sage UK: London, England}
}

@article{muller2012grounded,
  title={Grounded theory method in human-computer interaction and computer-supported cooperative work},
  author={Muller, Michael J and Kogan, Sandra},
  journal={The Human Computer Interaction Handbook (3 ed.), Julie A. Jacko (Ed.). CRC Press, Boca Raton, FL},
  pages={1003--1024},
  year={2012}
}

@article{friedman1937use,
  title={The use of ranks to avoid the assumption of normality implicit in the analysis of variance},
  author={Friedman, Milton},
  journal={Journal of the american statistical association},
  volume={32},
  number={200},
  pages={675--701},
  year={1937},
  publisher={Taylor \& Francis}
}

@techreport{conover1979multiple,
  title={Multiple-comparisons procedures},
  author={Conover, WJ and Iman, RL},
  year={1979},
  institution={Los Alamos National Laboratory (LANL)},
  address={Los Alamos, NM (United States)},
  number={LA-7677-MS},
  doi={10.2172/6057803},
  url={https://doi.org/10.2172/6057803}
}

@book{alvo2014statistical,
  title={Statistical methods for ranking data},
  author={Alvo, Mayer and Philip, LH},
  volume={1341},
  year={2014},
  publisher={Springer}
}

@article{harzing2009rating,
  title={Rating versus ranking: What is the best way to reduce response and language bias in cross-national research?},
  author={Harzing, Anne-Wil and Baldueza, Joyce and Barner-Rasmussen, Wilhelm and Barzantny, Cordula and Canabal, Anne and Davila, Anabella and Espejo, Alvaro and Ferreira, Rita and Giroud, Axele and Koester, Kathrin and others},
  journal={International business review},
  volume={18},
  number={4},
  pages={417--432},
  year={2009},
  publisher={Elsevier}
}

@article{yannakakis2015ratings,
  title={Ratings are overrated!},
  author={Yannakakis, Georgios N and Mart{\'\i}nez, H{\'e}ctor P},
  journal={Frontiers in ICT},
  volume={2},
  pages={13},
  year={2015},
  publisher={Frontiers Media SA}
}

@article{bloom1968learning,
  title={Learning for Mastery. Instruction and Curriculum. Regional Education Laboratory for the Carolinas and Virginia, Topical Papers and Reprints, Number 1.},
  author={Bloom, Benjamin S},
  journal={Evaluation comment},
  volume={1},
  number={2},
  pages={n2},
  year={1968},
  publisher={ERIC}
}

@article{krathwohl2002revision,
  title={A Revision Bloom's Taxonomy: An Overview},
  author={Krathwohl, DR},
  journal={Theory into Practice},
  year={2002}
}

@article{canas2004varieties,
  title={Varieties of concept mapping},
  author={Ca{\~n}as, AJ and Novak, JD and Gonz{\'a}lez, FM},
  year={2004},
  publisher={Citeseer},
  journal={Proceedings of the First International Conference on Concept Mapping}
}

@article{novak1990concept,
  title={Concept mapping: A useful tool for science education},
  author={Novak, Joseph D},
  journal={Journal of research in science teaching},
  volume={27},
  number={10},
  pages={937--949},
  year={1990},
  publisher={Wiley Online Library}
}

@article{delugach2016knowledge,
  title={A knowledge capture approach for directly acquiring team mental models},
  author={Delugach, Harry S and Etzkorn, Letha H and Carpenter, Sandra and Utley, Dawn},
  journal={International Journal of Human-Computer Studies},
  volume={96},
  pages={12--21},
  year={2016},
  publisher={Elsevier}
}

@inproceedings{moon2018case,
  title={A case for the superiority of concept mapping-based assessments for assessing mental models},
  author={Moon, Brian and Johnston, Charles and Moon, Skyler},
  booktitle={Concept Mapping: Renewing Learning and Thinking. Proceedings of the 8th Int. Conference on Concept Mapping, Medell{\'\i}n, Colombia: Universidad EAFIT},
  year={2018}
}

@article{o2004measuring,
  title={Measuring team cognition: Concept mapping elicitation as a means of constructing team shared mental models in an applied setting},
  author={O'Connor, Debra L and Johnson, Tristan E and Khalil, Mohammed K},
  year={2004},
  publisher={Universidad P{\'u}blica de Navarra},
  journal={Proceedings of the First International Conference on Concept Mapping}
}

@inproceedings{reese2004assessment,
  title={Assessment and concept map structure: Interaction between subscores and well-formed mental models},
  author={Reese, Debbie Denise},
  booktitle={meeting of the American Educational Research Association, San Diego},
  year={2004}
}

@inproceedings{doughty2024comparative,
  title={A comparative study of AI-generated (GPT-4) and human-crafted MCQs in programming education},
  author={Doughty, Jacob and Wan, Zipiao and Bompelli, Anishka and Qayum, Jubahed and Wang, Taozhi and Zhang, Juran and Zheng, Yujia and Doyle, Aidan and Sridhar, Pragnya and Agarwal, Arav and others},
  booktitle={Proceedings of the 26th Australasian Computing Education Conference},
  pages={114--123},
  year={2024}
}

@inproceedings{elkins2024teachers,
  title={How Teachers Can Use Large Language Models and Bloom’s Taxonomy to Create Educational Quizzes},
  author={Elkins, Sabina and Kochmar, Ekaterina and Cheung, Jackie CK and Serban, Iulian},
  booktitle={Proceedings of the AAAI Conference on Artificial Intelligence},
  volume={38},
  number={21},
  pages={23084--23091},
  year={2024}
}

@book{2023Introduc,
  title={Introduction to SQL and NoSQL Databases},
  isbn={9791192932002},
  year={2023},
  publisher={Saengneung Publishing},
  author={Park, Sungjin}
}

@book{carroll2017introduction,
  title={An introduction to modern astrophysics},
  author={Carroll, Bradley W and Ostlie, Dale A},
  year={2017},
  publisher={Cambridge University Press}
}

@book{2023sampling,
  title={Sampling Methodology},
  isbn={9791160736052},
  year={2023},
  publisher={Kyungmoon Publishing},
  author={Kim, Hoil}
}

@article{collective,
  title={Construction and Analysis of Collaborative Educational Networks based on Student Concept Maps},
  author={Freedman, Hayden and Young, Neil and Schaefer, David and Song, Qingyu and van der Hoek, Andr{\'e} and Tomlinson, Bill},
  journal={Proceedings of the ACM on Human-Computer Interaction},
  volume={8},
  number={CSCW1},
  pages={1--22},
  year={2024},
  publisher={ACM New York, NY, USA}
}

@inproceedings{kazemitabaar2024codeaid,
  title={Codeaid: Evaluating a classroom deployment of an llm-based programming assistant that balances student and educator needs},
  author={Kazemitabaar, Majeed and Ye, Runlong and Wang, Xiaoning and Henley, Austin Zachary and Denny, Paul and Craig, Michelle and Grossman, Tovi},
  booktitle={Proceedings of the CHI Conference on Human Factors in Computing Systems},
  pages={1--20},
  year={2024}
}

@inproceedings{lee2025impact,
  title={The impact of generative AI on critical thinking: Self-reported reductions in cognitive effort and confidence effects from a survey of knowledge workers},
  author={Lee, Hao-Ping and Sarkar, Advait and Tankelevitch, Lev and Drosos, Ian and Rintel, Sean and Banks, Richard and Wilson, Nicholas},
  booktitle={Proceedings of the 2025 CHI conference on human factors in computing systems},
  pages={1--22},
  year={2025}
}

@article{kosmyna2025your,
  title={Your brain on chatgpt: Accumulation of cognitive debt when using an ai assistant for essay writing task},
  author={Kosmyna, Nataliya and Hauptmann, Eugene and Yuan, Ye Tong and Situ, Jessica and Liao, Xian-Hao and Beresnitzky, Ashly Vivian and Braunstein, Iris and Maes, Pattie},
  journal={arXiv preprint arXiv:2506.08872},
  year={2025}
}

@inproceedings{chen2023gptutor,
  title={GPTutor: a ChatGPT-powered programming tool for code explanation},
  author={Chen, Eason and Huang, Ray and Chen, Han-Shin and Tseng, Yuen-Hsien and Li, Liang-Yi},
  booktitle={International conference on artificial intelligence in education},
  pages={321--327},
  year={2023},
  organization={Springer}
}

@article{phung2023generating,
  title={Generating high-precision feedback for programming syntax errors using large language models},
  author={Phung, Tung and Cambronero, Jos{\'e} and Gulwani, Sumit and Kohn, Tobias and Majumdar, Rupak and Singla, Adish and Soares, Gustavo},
  journal={arXiv preprint arXiv:2302.04662},
  year={2023}
}

@inproceedings{pal2024autotutor,
  title={Autotutor meets large language models: A language model tutor with rich pedagogy and guardrails},
  author={Pal Chowdhury, Sankalan and Zouhar, Vil{\'e}m and Sachan, Mrinmaya},
  booktitle={Proceedings of the Eleventh ACM Conference on Learning@ Scale},
  pages={5--15},
  year={2024}
}

@inproceedings{draxler2023relevance,
  title={Relevance, effort, and perceived quality: Language learners’ experiences with AI-generated contextually personalized learning material},
  author={Draxler, Fiona and Schmidt, Albrecht and Chuang, Lewis L},
  booktitle={Proceedings of the 2023 ACM Designing Interactive Systems Conference},
  pages={2249--2262},
  year={2023}
}

@inproceedings{leong2024putting,
  title={Putting things into context: Generative AI-enabled context personalization for vocabulary learning improves learning motivation},
  author={Leong, Joanne and Pataranutaporn, Pat and Danry, Valdemar and Perteneder, Florian and Mao, Yaoli and Maes, Pattie},
  booktitle={Proceedings of the 2024 CHI Conference on Human Factors in Computing Systems},
  pages={1--15},
  year={2024}
}

@article{pardos2024chatgpt,
  title={ChatGPT-generated help produces learning gains equivalent to human tutor-authored help on mathematics skills},
  author={Pardos, Zachary A and Bhandari, Shreya},
  journal={Plos one},
  volume={19},
  number={5},
  pages={e0304013},
  year={2024},
  publisher={Public Library of Science San Francisco, CA USA}
}

@article{bastani2024generative,
  title={Generative ai can harm learning},
  author={Bastani, Hamsa and Bastani, Osbert and Sungu, Alp and Ge, Haosen and Kabakc{\i}, Ozge and Mariman, Rei},
  journal={Available at SSRN},
  volume={4895486},
  year={2024}
}

@article{peters2024ai,
  title={AI and the future of humanity: ChatGPT-4, philosophy and education--Critical responses},
  author={Peters, Michael A and Jackson, Liz and Papastephanou, Marianna and Jandri{\'c}, Petar and Lazaroiu, George and Evers, Colin W and Cope, Bill and Kalantzis, Mary and Araya, Daniel and Tesar, Marek and others},
  journal={Educational Philosophy and Theory},
  volume={56},
  number={9},
  pages={828--862},
  year={2024},
  publisher={Taylor \& Francis}
}

@article{dwivedi2023opinion,
  title={Opinion Paper:“So what if ChatGPT wrote it?” Multidisciplinary perspectives on opportunities, challenges and implications of generative conversational AI for research, practice and policy},
  author={Dwivedi, Yogesh K and Kshetri, Nir and Hughes, Laurie and Slade, Emma Louise and Jeyaraj, Anand and Kar, Arpan Kumar and Baabdullah, Abdullah M and Koohang, Alex and Raghavan, Vishnupriya and Ahuja, Manju and others},
  journal={International journal of information management},
  volume={71},
  pages={102642},
  year={2023},
  publisher={Elsevier}
}

@article{mogavi2024chatgpt,
  title={ChatGPT in education: A blessing or a curse? A qualitative study exploring early adopters’ utilization and perceptions},
  author={Mogavi, Reza Hadi and Deng, Chao and Kim, Justin Juho and Zhou, Pengyuan and Kwon, Young D and Metwally, Ahmed Hosny Saleh and Tlili, Ahmed and Bassanelli, Simone and Bucchiarone, Antonio and Gujar, Sujit and others},
  journal={Computers in Human Behavior: Artificial Humans},
  volume={2},
  number={1},
  pages={100027},
  year={2024},
  publisher={Elsevier}
}

@inproceedings{weidinger2022taxonomy,
  title={Taxonomy of risks posed by language models},
  author={Weidinger, Laura and Uesato, Jonathan and Rauh, Maribeth and Griffin, Conor and Huang, Po-Sen and Mellor, John and Glaese, Amelia and Cheng, Myra and Balle, Borja and Kasirzadeh, Atoosa and others},
  booktitle={Proceedings of the 2022 ACM conference on fairness, accountability, and transparency},
  pages={214--229},
  year={2022}
}

@article{sage2019reading,
  title={Reading from print, computer, and tablet: Equivalent learning in the digital age},
  author={Sage, Kara and Augustine, Heather and Shand, Hannah and Bakner, Kaelah and Rayne, Sidny},
  journal={Education and Information Technologies},
  volume={24},
  number={4},
  pages={2477--2502},
  year={2019},
  publisher={Springer}
}

@article{rockinson2013electronic,
  title={Electronic versus traditional print textbooks: A comparison study on the influence of university students' learning},
  author={Rockinson-Szapkiw, Amanda J and Courduff, Jennifer and Carter, Kimberly and Bennett, David},
  journal={Computers \& Education},
  volume={63},
  pages={259--266},
  year={2013},
  publisher={Elsevier}
}

@article{edyburn1991fact,
  title={Fact Retrieval by students with and without learning handicaps using print and electronic encyclopedias},
  author={Edyburn, Dave L},
  journal={Journal of Special Education Technology},
  volume={11},
  number={2},
  pages={75--90},
  year={1991},
  publisher={SAGE Publications Sage CA: Los Angeles, CA}
}

@article{eppler2004concept,
  title={The concept of information overload: A review of literature from organization science, accounting, marketing, MIS, and related disciplines},
  author={Eppler, Martin J and Mengis, Jeanne},
  journal={The information society},
  volume={20},
  number={5},
  pages={325--344},
  year={2004},
  publisher={Taylor \& Francis}
}

@article{thompson2013digital,
  title={The digital natives as learners: Technology use patterns and approaches to learning},
  author={Thompson, Penny},
  journal={Computers \& Education},
  volume={65},
  pages={12--33},
  year={2013},
  publisher={Elsevier}
}

@article{walraven2008information,
  title={Information-problem solving: A review of problems students encounter and instructional solutions},
  author={Walraven, Amber and Brand-Gruwel, Saskia and Boshuizen, Henny PA},
  journal={Computers in Human Behavior},
  volume={24},
  number={3},
  pages={623--648},
  year={2008},
  publisher={Elsevier}
}

@inproceedings{avula2018searchbots,
  title={Searchbots: User engagement with chatbots during collaborative search},
  author={Avula, Sandeep and Chadwick, Gordon and Arguello, Jaime and Capra, Robert},
  booktitle={Proceedings of the 2018 conference on human information interaction \& retrieval},
  pages={52--61},
  year={2018}
}

@article{perez2020rediscovering,
  title={Rediscovering the use of chatbots in education: A systematic literature review},
  author={P{\'e}rez, Jos{\'e} Quiroga and Daradoumis, Thanasis and Puig, Joan Manuel Marqu{\`e}s},
  journal={Computer Applications in Engineering Education},
  volume={28},
  number={6},
  pages={1549--1565},
  year={2020},
  publisher={Wiley Online Library}
}

@article{vakkari2016searching,
  title={Searching as learning: A systematization based on literature},
  author={Vakkari, Pertti},
  journal={Journal of Information Science},
  volume={42},
  number={1},
  pages={7--18},
  year={2016},
  publisher={SAGE Publications Sage UK: London, England}
}

@article{jansen2009using,
  title={Using the taxonomy of cognitive learning to model online searching},
  author={Jansen, Bernard J and Booth, Danielle and Smith, Brian},
  journal={Information Processing \& Management},
  volume={45},
  number={6},
  pages={643--663},
  year={2009},
  publisher={Elsevier}
}

@article{weyer1982design,
  title={The design of a dynamic book for information search},
  author={Weyer, Stephen A},
  journal={International Journal of Man-Machine Studies},
  volume={17},
  number={1},
  pages={87--107},
  year={1982},
  publisher={Elsevier}
}

@inproceedings{kim2012understanding,
  title={Understanding book search behavior on the web},
  author={Kim, Jin Young and Feild, Henry and Cartright, Marc},
  booktitle={Proceedings of the 21st ACM international conference on Information and knowledge management},
  pages={744--753},
  year={2012}
}

@article{r2006throughput,
  title={Throughput: a simple performance index with desirable characteristics},
  author={R. Thorne, David},
  journal={Behavior research methods},
  volume={38},
  number={4},
  pages={569--573},
  year={2006},
  publisher={Springer}
}

@inproceedings{mo2025conversational,
  title={Conversational search: From fundamentals to frontiers in the LLM era},
  author={Mo, Fengran and Meng, Chuan and Aliannejadi, Mohammad and Nie, Jian-Yun},
  booktitle={Proceedings of the 48th International ACM SIGIR Conference on Research and Development in Information Retrieval},
  pages={4094--4097},
  year={2025}
}

@article{paas1993efficiency,
  title={The efficiency of instructional conditions: An approach to combine mental effort and performance measures},
  author={Paas, Fred GWC and Van Merri{\"e}nboer, Jeroen JG},
  journal={Human factors},
  volume={35},
  number={4},
  pages={737--743},
  year={1993},
  publisher={SAGE Publications Sage CA: Los Angeles, CA}
}

@article{kiewra1989review,
  title={A review of note-taking: The encoding-storage paradigm and beyond},
  author={Kiewra, Kenneth A},
  journal={Educational Psychology Review},
  volume={1},
  number={2},
  pages={147--172},
  year={1989},
  publisher={Springer}
}

@article{chen2025more,
  title={More ai assistance reduces cognitive engagement: Examining the ai assistance dilemma in ai-supported note-taking},
  author={Chen, Xinyue and Ruan, Kunlin and Ju, Kexin Phyllis and Yap, Nathan and Wang, Xu},
  journal={Proceedings of the ACM on Human-Computer Interaction},
  volume={9},
  number={7},
  pages={1--29},
  year={2025},
  publisher={ACM New York, NY, USA}
}

@book{estes2022handbook,
  title={Handbook of learning and cognitive processes},
  author={Estes, William},
  year={2022},
  publisher={Psychology Press}
}

@book{simon1978information,
  title={Information-processing theory of human problem solving},
  author={Simon, Herbert A},
  year={1978},
  publisher={Erlbaum Hillsdale, NJ}
}

@book{feigenbaum1959information,
  title={An information processing theory of verbal learning},
  author={Feigenbaum, Edward A},
  year={1959},
  publisher={Rand Corporation}
}

@book{lindsay2013human,
  title={Human information processing: An introduction to psychology},
  author={Lindsay, Peter H and Norman, Donald A},
  year={2013},
  publisher={Academic press}
}

@article{gao2023coaicoder,
  title={CoAIcoder: Examining the effectiveness of AI-assisted human-to-human collaboration in qualitative analysis},
  author={Gao, Jie and Choo, Kenny Tsu Wei and Cao, Junming and Lee, Roy Ka-Wei and Perrault, Simon},
  journal={ACM Transactions on Computer-Human Interaction},
  volume={31},
  number={1},
  pages={1--38},
  year={2023},
  publisher={ACM New York, NY}
}

@inproceedings{shaer2024ai,
  title={AI-Augmented Brainwriting: Investigating the use of LLMs in group ideation},
  author={Shaer, Orit and Cooper, Angelora and Mokryn, Osnat and Kun, Andrew L and Ben Shoshan, Hagit},
  booktitle={Proceedings of the 2024 CHI Conference on Human Factors in Computing Systems},
  pages={1--17},
  year={2024}
}

@inproceedings{dhillon2024shaping,
  title={Shaping human-AI collaboration: Varied scaffolding levels in co-writing with language models},
  author={Dhillon, Paramveer S and Molaei, Somayeh and Li, Jiaqi and Golub, Maximilian and Zheng, Shaochun and Robert, Lionel Peter},
  booktitle={Proceedings of the 2024 CHI Conference on Human Factors in Computing Systems},
  pages={1--18},
  year={2024}
}

@article{joiner2008long,
  title={Long-term retention explained by a model of short-term learning in the adaptive control of reaching},
  author={Joiner, Wilsaan M and Smith, Maurice A},
  journal={Journal of neurophysiology},
  volume={100},
  number={5},
  pages={2948--2955},
  year={2008},
  publisher={American Physiological Society}
}

@article{Miller2025WorkloadGap,
  title        = {I Ran The Numbers. There is a 300\% Workload Gap Between Some Majors.},
  author       = {Jacob M. Miller},
  journal      = {The Harvard Crimson (Opinion)},
  year         = {2025},
  month        = {Feb 27},
  url          = {https://www.thecrimson.com/article/2025/2/27/miller-harvard-course-workload-divisions/}
}

@incollection{atkinson1968human,
  title={Human memory: A proposed system and its control processes},
  author={Atkinson, Richard C and Shiffrin, Richard M},
  booktitle={Psychology of learning and motivation},
  volume={2},
  pages={89--195},
  year={1968},
  publisher={Elsevier}
}

@article{besterfield2004scoring,
  title={Scoring concept maps: An integrated rubric for assessing engineering education},
  author={Besterfield-Sacre, Mary and Gerchak, Jessica and Lyons, Mary Rose and Shuman, Larry J and Wolfe, Harvey},
  journal={Journal of Engineering Education},
  volume={93},
  number={2},
  pages={105--115},
  year={2004},
  publisher={Wiley Online Library}
}

@article{watson2016assessing,
  title={Assessing conceptual knowledge using three concept map scoring methods},
  author={Watson, Mary Katherine and Pelkey, Joshua and Noyes, Caroline R and Rodgers, Michael O},
  journal={Journal of engineering education},
  volume={105},
  number={1},
  pages={118--146},
  year={2016},
  publisher={Wiley Online Library}
}

@article{kintsch1988burden,
  title        = {The role of knowledge in discourse comprehension: A construction-integration model.},
  author       = {Walter Kintsch},
  journal      = {Psychological Review},
  volume={95},
  number={2},
  pages={163--182},
  year         = {1988},
  url          = {https://doi.org/10.1037/0033-295X.95.2.163}
}

@article{sweller1998burden,
  title={Cognitive architecture and instructional design},
  author={Sweller, John and Van Merrienboer, Jeroen JG and Paas, Fred GWC},
  journal={Educational psychology review},
  volume={10},
  number={3},
  pages={251--296},
  year={1998},
  publisher={Springer}
}

@article{cromley2024three,
  title={Three applications of semantic network analysis to individual student think-aloud data},
  author={Cromley, Jennifer G and Mirabelli, Joseph F and Kunze, Andrea J},
  journal={Contemporary Educational Psychology},
  volume={79},
  pages={102318},
  year={2024},
  publisher={Elsevier}
}

@article{bjork2011making,
  title={Making things hard on yourself, but in a good way: Creating desirable difficulties to enhance learning},
  author={Bjork, Elizabeth L and Bjork, Robert A and others},
  journal={Psychology and the real world: Essays illustrating fundamental contributions to society},
  volume={2},
  number={59-68},
  pages={56--64},
  year={2011}
}

@inproceedings{rowlands2008google,
  title={The Google generation: the information behaviour of the researcher of the future},
  author={Rowlands, Ian and Nicholas, David and Williams, Peter and Huntington, Paul and Fieldhouse, Maggie and Gunter, Barrie and Withey, Richard and Jamali, Hamid R and Dobrowolski, Tom and Tenopir, Carol},
  booktitle={Aslib proceedings},
  volume={60},
  number={4},
  pages={290--310},
  year={2008},
  organization={Emerald Group Publishing Limited}
}

@book{brabazon2016university,
  title={The University of Google: Education in the (post) information age},
  author={Brabazon, Tara},
  year={2016},
  publisher={Routledge}
}

@incollection{sweller2011cognitive,
  title={Cognitive load theory},
  author={Sweller, John},
  booktitle={Psychology of learning and motivation},
  volume={55},
  pages={37--76},
  year={2011},
  publisher={Elsevier}
}

@article{graham2006blended,
  title={Blended learning systems: Definition, current trends, and future directions},
  author={Graham, Charles R and others},
  journal={The handbook of blended learning: Global perspectives, local designs},
  volume={1},
  pages={3--21},
  year={2006}
}

@article{akccayir2018flipped,
  title={The flipped classroom: A review of its advantages and challenges},
  author={Ak{\c{c}}ay{\i}r, G{\"o}k{\c{c}}e and Ak{\c{c}}ay{\i}r, Murat},
  journal={Computers \& Education},
  volume={126},
  pages={334--345},
  year={2018},
  publisher={Elsevier}
}

\appendix
\newpage
\section{MCQ GENERATION}
\label{sec:MCQgeneration}

Multiple-choice questions (MCQs) were developed to evaluate the learning outcomes of the experiment participants. A total of 27 questions were created, with 9 questions for each STEM module, consisting of 3 questions from each of the selected taxonomy categories. These questions were designed to cover each module's learning objectives (LO). According to previous studies~\cite{elkins2024teachers,doughty2024comparative}, generating MCQs through large language models (LLMs) is preferable due to its time and cost efficiency, quality comparable to human-generated items, and alignment with Bloom's taxonomy; therefore, the MCQs were generated using ChatGPT-4o~\cite{chatgpt4o}. 

\subsection{Prompts}

\begin{figure}[h!]
    \centerline{\includegraphics[width=\columnwidth]{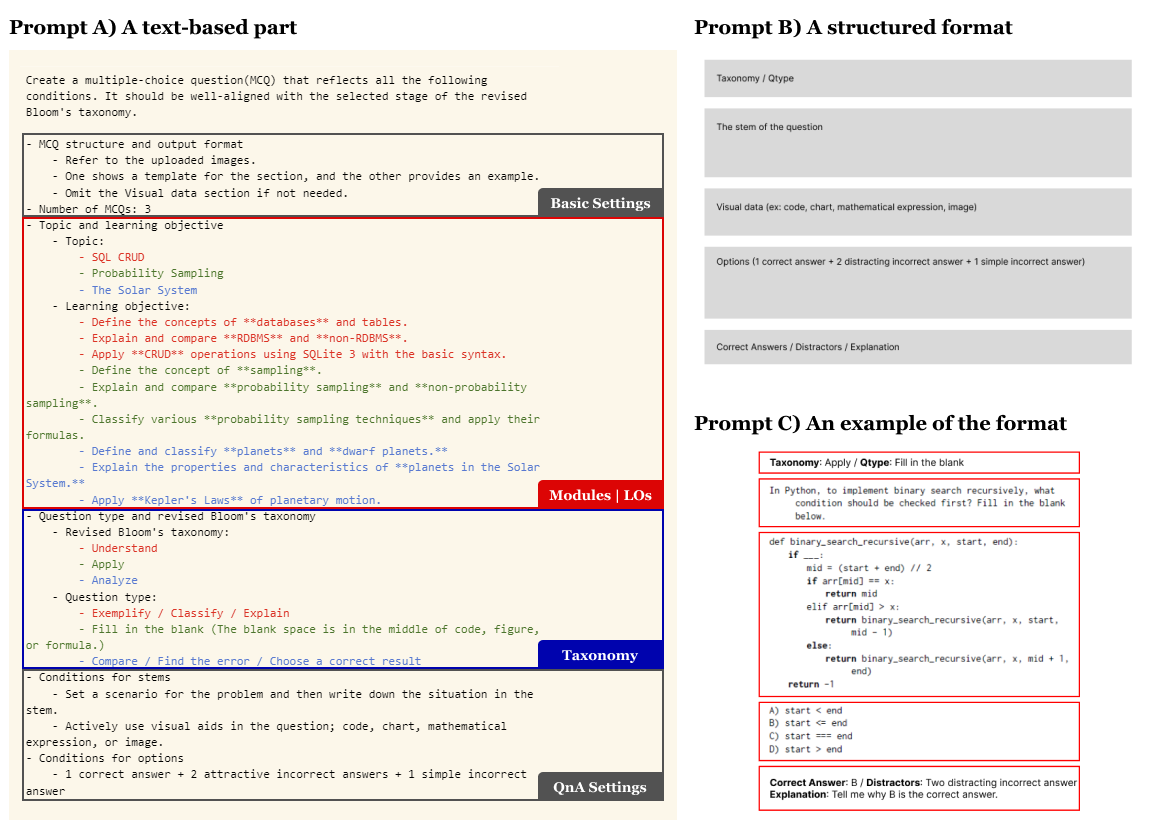}}
    \caption{Text (A) and images (B, C) were provided together to GPT-4. For the \textbf{Modules | LOs} and \textbf{Taxonomy} sections, only one color was retained for each to match their respective purposes.}
    \Description{The figure presents three prompts provided to GPT-4 for MCQ generation. Prompt A is a text-based prompt detailing the structure and requirements for creating MCQs, including basic settings, learning objectives, and taxonomy alignment. Prompt B shows a structured format template used to ensure consistency across questions, including taxonomy type, question structure, and answer choices. Prompt C provides an example of the format, highlighting a taxonomy-aligned question. For the Modules | LOs and Taxonomy sections, only one color was retained to align with their specific purposes.}
    \label{fig:prompt_figure}
\end{figure}

As shown in \autoref{fig:prompt_figure}, a text-based prompt specified the question requirements. To ensure consistent formatting, we also included images showing the exact format guidelines.
\vspace{-1em}

\subsection{Quality Evaluation}

The evaluation was conducted using two levels of metrics: question-level~\cite{doughty2024comparative} and quiz-level~\cite{elkins2024teachers,doughty2024comparative}.
We initially generated three times more items than required to allow quality screening. In total, 81 MCQs were created (27 per module), and the closed-book test containing 9 MCQs per module was finalized through a three-step refinement process. 
\textit{Step 1)} Four of the authors cross-validated 81 initial questions using metrics (\autoref{tab:question-level}), and two of them further reviewed the questions for alignment with learning objectives (LO) and taxonomy, removing items that did not meet the criteria. 
\textit{Step 2)} After refinement and evaluation, the highest-rated items based on the question-level metrics were selected, resulting in a closed-book test consisting of 9 MCQs per module.
\textit{Step 3)} For the quiz-level evaluation, we recruited three domain experts, all of whom were university professors. The experts evaluated the closed-book tests based on three criteria---structure, redundancy, and usefulness (\autoref{tab:expert_eval}).
\idx{7}\revised{Experts received 100,000 KRW (approx.73 USD) as compensation for participation.}


\begin{table}[h!]
    \centering
    \small
    \begin{tabular}{c|c|m{4cm}|m{6cm}} \hline\hline

         &  \textbf{Rubric item}& \centering  \textbf{Question}& \multicolumn{1}{c}{\textbf{Options}}\\ \hline\hline

         M1&  Fluency&  Is the language grammatically correct and clear?& (1) Yes, it is written in grammatically correct and clear language.

(2) No, it is not written in grammatically correct and clear language.

(3) I am unsure.\\ \hline 

         M2&  Correct answer&  Does the correct answer appear within the choices? If so, is the option marked as correct the right answer?& (1) Yes, the correct answer is present, and the option is marked as the 'correct' answer.

(2) The correct answer is present, but it is not marked as the 'correct' option.

(3) There are multiple correct answers.

(4) No, the correct answer is not present among the options.

(5) I am unsure.\\ \hline 

         M3&  Unique choices&  Are the answer choices distinct and unique from one another?& (1) Yes, the answer choices are completely distinct from one another.

(2) Some choices are distinct, but others are too similar.

(3) No, they all seem similar and appear to overlap.

(4) I am unsure.\\ \hline 

         M4&  No obviously wrong choice&  Are there any answer choices that are incorrect or wrong?& (1) Yes, there are no incorrect answer choices.

(2) Yes, but the correct answer is too easy to infer.

(3) No, there are incorrect choices.

(4) I am unsure\\ \hline 

         M5&  Correct material&  If supplementary materials (e.g., code, formulas, images) are included in the question or choices, do they make sense grammatically and logically?& (1) Yes, the supplementary materials are grammatically and logically well-constructed.

(2) There are minor issues.

(3) No, the materials are incomprehensible.

(4) I am unsure.\\ \hline 

         M6&  LO alignment&  Does this question contribute to achieving the learning objectives?& (1) Yes, it contributes to achieving the learning objectives.

(2) It probably does, but there are significant gaps.

(3) No, it does not help achieve the learning objectives.

(4) I am unsure.\\ \hline 

         M7&  Taxonomy alignment&  Is the question appropriately aligned with the intended Bloom's taxonomy level?& (1) Yes, the question is aligned with the intended taxonomy.

(2) No, the question is unrelated to the intended taxonomy.

(3) I am unsure.\\ \hline\hline

    \end{tabular}

    \caption{The question-level metric was used to evaluate the appropriateness of the initial 81 questions and additional generated questions, resulting in the selection of 27 questions. Each question underwent cross-evaluation by at least three authors.}

    \Description{The rubric evaluates multiple-choice questions (MCQs) on seven criteria: fluency, correct answer, unique choices, absence of obviously wrong choices, correctness of supplementary materials, alignment with learning objectives, and alignment with Bloom’s taxonomy. Each criterion is assessed through a set of options that range from "yes" or "no" to specific conditions such as whether multiple correct answers are present or if supplementary materials are comprehensible.}

    \label{tab:question-level}

\end{table}

\subsection{Examples of MCQs}
\begin{figure}[h]
    \centerline{\includegraphics[width=\columnwidth]{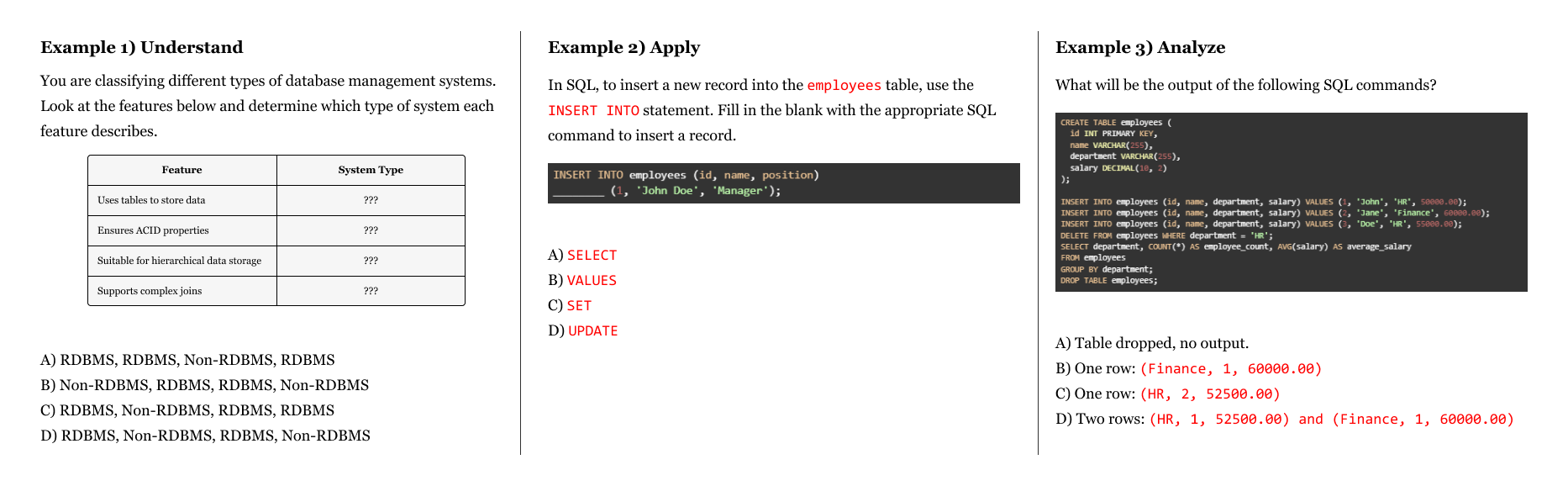}}
    \caption{As part of the post-test, nine questions from the database module were used, and one example from each taxonomy is provided as follows.}
    \Description{Three post-test questions from the database module, categorized by Bloom's taxonomy: 1) "Understand" asks participants to classify database systems based on features. 2) "Apply" asks for the correct SQL command to insert data into a table. 3) "Analyze" requires determining the output of a given SQL script.}
    \label{fig:MCQ_ex}
\end{figure}

\end{document}